\documentclass[12pt,a4paper]{article}
\usepackage[dvips]{graphicx}
\usepackage[english]{babel}
\usepackage{multirow}
\usepackage{amsmath}
\usepackage{amssymb}
\usepackage{rotating}
\usepackage{latexsym}
\usepackage[dvips]{color}
\usepackage{subfigure}
\usepackage{booktabs}
\usepackage{caption}
\usepackage{subfigure}
\usepackage{booktabs}
\usepackage{caption}
\usepackage{setspace}
\usepackage{hyperref}
\usepackage{array}
\usepackage{xmpmulti}
\usepackage[hmargin=2.5cm,vmargin=2.5cm]{geometry}

\begin{document}

\title{Option pricing with non-Gaussian scaling and infinite-state switching volatility}
\author{Fulvio Baldovin\thanks{University of Padova, Department of Physics and Astronomy} \and Massimiliano Caporin\thanks{University of Padova, Department of Economics and Management, corresponding author: massimiliano.caporin@unipd.it} \and Michele Caraglio\footnotemark[1] \and Attilio L. Stella\footnotemark[1] \and Marco Zamparo\thanks{Politecnico di Torino, Department of Applied Science and Technology, and Human Genetics Foundation Torino}}
\date{}
\maketitle

\noindent \textbf{Abstract:} Volatility clustering, long-range dependence, and non-Gaussian scaling are stylized facts of financial assets dynamics. They are ignored in the Black \& Scholes framework, but have a relevant impact on the pricing of options written on financial assets. Using a recent model for market dynamics which adequately captures the above stylized facts, we derive closed form equations for option pricing, obtaining the Black \& Scholes as a special case. By applying our pricing equations to a major equity index option dataset, we show that inclusion of stylized features in financial modeling moves derivative prices about $30\%$ closer to the market values without the need of calibrating models parameters on available derivative prices.


\doublespacing

\section{Introduction}

Dynamical properties and stylized facts of financial time series
have raised considerable interest in both theoretical and applied
econometrics. Within these fields, one topic widely discussed 
is related to the memory properties observed over return sequences, in
particular regarding their absolute value (volatility). 
A variety of techniques have been developed to
include realistic volatility dynamic features into a continuous time model, thus improving
the geometric Brownian motion assumption underlying the classic Black-Scholes (BS)
model. We mention, among many others, the stochastic volatility, see, e.g., Fouque et al. (2000) and therein cited
references, the and GARCH-based approach of Heston (1993), and Heston and Nandi (2000).
In what follows, volatility memory properties, and in particular the long-memory behaviour, are taken into account on the
basis of the ``scaling symmetry'' concept, that has a long
tradition in statistical mechanics and in the 
physics of complex systems. Our main purpose is to discuss the impact of memory
and scaling properties on the price of financial
derivatives, and to compare them with those coming from the BS framework. 

Long-memory
behaviour has been often associated with the evaluation of the
Hurst exponent, $H$, introduced in the seminal work of Hurst
(1951).
The Hurst exponent links 
the time series fluctuations with the time scale over which they are observed. Indeed, in the presence of a simple scaling 
symmetry for a stochastic process $X_t$ ($t=1,2,\dots$)
supposed to generate the 
time series of interest, the 
$q-$th order moment satisfies
$\mathbb{E}\left[\vert \sum_{t=1}^\tau X_{t} \vert ^q \right]
\sim A_q \tau ^{qH}$, 
where $\tau$ defines the time scale at which the
terms of the series are aggregated and
$A_q$ is an amplitude. This follows from the fact that
the probability density function of $X_{1}+X_{2}+\ldots +X_{\tau}$
satisfies the scaling identity, $\tau ^H g_\tau(\tau ^H\ x)=g(x)$,
where $g(\cdot)$ is the scaling function and $H$ is also called scaling exponent.
Well-studied forms of
scaling were first observed in the study of critical phenomena, as in a critical magnetic system, see 
(see, e.g., 
Sethna et al. (2001), and therein cited references, or disordered systems, 
see Bouchaud and Georges (1990).
Indeed, if the elements of a time series are extracted independently
from a Gaussian probability density function,
the Hurst exponent $H$ has a value
equal to $1/2$. In several empirical studies, deviations from
this reference behaviour have been observed, and are referred to as forms of
``anomalous scaling''. These deviations are
generally considered a consequence of long-memory effects.\footnote{However, 
we point out that, strictly speaking, an anomalous Hurst exponent 
is not necessarily a consequence to the
existence of long-memory, Cont (2005).} 
It is important to point out that
anomalous scaling conditions can also arise in cases in which the
scaling exponent is equal to $1/2$, whenever $g$ is not Gaussian. 
Finally, a more general form of anomalous scaling 
called ``multiscaling'' is
associated with those cases in which the scaling exponent depends on
the moment order $q$. If multiscaling is present, the scaling or Hurst
exponent $H$ can be expressed as a function of the moment order,
leading to $H=H(q)$. 
Non-Gaussian scaling plays a central role in the present study, where an interesting
case is made for the existence of multiscaling associated with
long-memory behaviour of financial data.

A seminal contribution linking scaling properties to the financial
framework was made by Mandelbrot (1963). Further studies
appeared only at the beginning of the '90s, mainly starting with the
newly available high frequency data on currencies. We cite, among
others, Muller et al. (1990), Dacorogna et al. (1993, 2001), and
Guillaume et
al. (1997). Subsequently,
additional works analysed the scaling properties of financial assets
other than currencies such as bonds (Ballocchi et al. 1999) and
equities (Di Matteo et al. 2005). Finally, scaling
properties of financial data have been also analysed from an
econo-physics point of view (see Mantegna and Stanley 1995, 
Ghashghaie et al. 1996, and Stanley and Plerou 2001), establishing  a connection with the physics of critical phenomena. 

The presence of non-Gaussian or anomalous scaling in financial log-return time series introduces
a relevant challenge for practitioners, in particular when their
purpose is the determination of a derivative contract price. 
As a matter of fact, the
dynamics of underlying prices assumed in the BS
framework is well known to be
not consistent with the scaling properties of equity indices. This
is a source of derivative pricing inefficiencies.

Rather than relying on concepts
like implied volatility to correct the BS framework,
our aim here is an evaluation of the effects on pricing of
a substantial improvement of the model describing the dynamics of
the underlying asset, while maintaining a link with the BS formalism.
The main contribution of the present paper is the derivation of closed-form formulae for 
option pricing and associated hedging strategy, based on an 
equivalent martingale measure.
This is obtained by means of a model recently introduced in statistical mechanics
by Zamparo et al. (2013) whose advantage is to be consistent with the observed features of financial 
data, thus including non-Gaussian scaling behaviour.
The general results provided here include as a special case the
BS pricing formula. 
Although we rely on a Monte Carlo computation to evaluate 
probabilities related to the underlying dynamics, 
these formulae define {\it a priori} 
the option price and the associated strategy, conditioned
on the previously observed values of the underlying asset's returns.
Also, we differ from the stochastic volatility pricing approaches 
(see, e.g., Fouque et al. 2000, or Heston, 1993) since our 
formulae are closed in the sense that the pricing parameters are
directly inferred from the underlying model dynamics
and do not need any calibration with respect to known derivatives'
market prices.
Indeed, an empirical application shows that compared with  
BS the proposed approach determines a 
price sensibly closer to the market one.
We postpone to future works the derivation of Greeks and the use of our pricing approach within option
trading strategies.

The paper proceeds as follows: Section 2 introduces the stochastic process 
for the underlying risky asset; Section 3 
reports the novel approach for option pricing based on the model presented in 
Section 2; an empirical analysis is outlined in Section 4 and Section 5
concludes. Proofs are reported in the Appendix.

\section{Non-Gaussian scaling in financial data}
\label{sec_model}
We consider the dynamics of a financial asset or index whose value at
time $t$ is denoted by $S_t$, $t=0,1,2,\dots$, while its log-returns
are given as $X_t=\log \left(S_t\right) - \log \left(S_{t-1}\right)$.
The existence of non-Gaussian scaling properties 
may be associated with different effects, including various kind of
data seasonality, see Dacorogna et al. (1993), among others, or long-range dependencies.
The latter have been extensively studied in statistics (Beran, 1994), econometrics
(Robinson, 2003), and finance by the pioneering work of Lo (1991). For
financial time series, and for time scales ranging from minutes to
months, significant deviations from Gaussianity of the scaling
function $g$ have been observed; see Cont (2001), among others.
Moreover, whereas a scaling exponent $H\simeq1/2$ is observed for
$q\lesssim 3$, multiscaling behaviour with $q$-dependent $H(q)<1/2$ is
often observed for higher-order moments (see, e.g., Di Matteo et al.,
2005).

Following the original suggestion by Baldovin and
Stella (2007), Zamparo et al. (2013) assume the anomalous scaling 
symmetry of the density for the aggregated returns. Taking this as a guiding criterion,
they propose stochastic processes whose realizations are able to 
replicate the long-range dependence and non-Gaussianity typical of a financial 
time series. At the same time, the use of scaling symmetry leads to processes 
which are relatively easy to manipulate for both calibration purposes and analytical 
calculations. High frequency models based on the same criterion were 
already proposed in Baldovin et al. (2011) and adopted for trading 
strategies in Baldovin et al. (2014). 

While referring to the original publication (Zamparo et al. 2013) for
motivations and details of the model construction, we give here the basic
definitions of the process used then in a derivative 
pricing framework. In the proposed structure, two independent stochastic components 
affect the asset returns' dynamical evolution.  The first one, is a fully {\it endogenous} process called $Y_t$, dependent on the
past asset 
history through an auto-regressive scheme of order $M$. 
Once the memory range $M$ is established,\footnote{The memory range $M$ can be fixed according to the data frequency, or with respect the objective of the researcher. With daily data, one might take $M=21$, that is one month, given it is a sufficient range to approximate a truly long-memory process as shown in Corsi (2009). Alternatively, one could set $M=60$, close to three months, to provide a better approximation.} the conditional probability
density function of $Y_t$ only depends on the previous $M$ values of
$Y_t$, as can be deduced by the following definitions for its joint
probability density $f_t^Y$:
\begin{equation}
  f_t^Y(y_1,\ldots,y_t)
\equiv\varphi_{1,t}(y_1,\ldots,y_t)
\label{gp}
\end{equation} 
if $t=1,2,\ldots,M+1$, and
\begin{eqnarray}
f_t^Y(y_1,\ldots,y_t)
&\equiv&\frac{\varphi_{t-M,t}(y_{t-M},\ldots,y_t)}{\varphi_{t-M,t-1}(y_{t-M},\ldots,y_{t-1})}\;
f_t^Y(y_1,\ldots,y_{t-1})
\label{gs}
\end{eqnarray} 
if $t>M+1$. The
probability densities $\varphi_t$ (for simplicity, we now suppress the first index) are assumed to be Gaussian mixtures:
\begin{equation}
\varphi_t(y_1,\ldots,y_t)
\equiv
\int_{0}^\infty \left(\prod_{i=1}^t \;\mathcal N_\sigma(y_i)\right) \;\rho(\sigma) \; d\sigma,
\label{gp_1}
\end{equation}
where $\mathcal N_\sigma$ is a Gaussian with volatility $\sigma$ and $\rho(\sigma) \geq 0$,
with $\int_0^{+\infty} \rho(\sigma) d\sigma=1$.
The $\varphi_t$ defined in (\ref{gp_1}) 
allows recovering proper empirical scaling properties 
while fat tailed densities for $\rho$ lead
to fat tailed marginal distributions for the $Y_t$
(Zamparo et al. 2013). Notably, the process defined in 
this way is similar to an ARCH process of order
$M$, with the difference that $Y_t$ is in general
heterodistributed in place of just heteroskedastic. Indeed, 
not only the variance, as in ordinary ARCH schemes, but also the form of 
the conditional PDF of $Y_t$ depends on the
previous $M$ values $y_{t-1},\ldots,y_{t-M}$. Specific
choices for the density $\rho$ give the advantage of enabling the
explicit integration over $\sigma$. This is what happens, 
for instance, when choosing $\sigma^2$ 
to be distributed according to an inverse-Gamma density. The
density $\rho\left(\sigma\right)$ is then associated with a
``shape parameter'' $\alpha$ and a ``scale parameter'' $\beta$,
according to 
\begin{equation}
\rho_{\alpha,\beta}(\sigma)=\frac{2^{1-\frac{\alpha}{2}}}{\Gamma(\frac{\alpha}{2})}
\;\frac{\beta^\alpha}{\sigma^{\alpha+1}}
\;\text{e}^{-\frac{\beta^2}{2\sigma^2}}.
\label{rhoinv}
\end{equation}
Upon explicit integration over $\sigma$, 
the endogenous component becomes in this case a genuine ARCH process (see Zamparo
et al., 2013) described by
\begin{equation}
Y_t=\begin{cases}
\beta\cdot Z_1 & \mbox{if } t=1;\\
\sqrt{\beta^2+\sum_{n=1}^{\min\{t-1,M\}}Y_{t-n}^2\;}\cdot Z_t & \mbox{if } t>1,
\end{cases}
\end{equation}
where the return residual process $\{Z_t\}_{t=1}^{\infty}$ is given as a sequence of
independent Student's t-distributed variables:
\begin{equation}
f_t^Z(z_1,\ldots,z_t)=\prod_{n=1}^t\frac{\Gamma(\frac{\alpha_n+1}{2})}{\sqrt{\pi}\;\Gamma(\frac{\alpha_n}{2})}
\;(1+z_n^2)^{-\frac{\alpha_n+1}{2}},
\label{Zdist}
\end{equation}
with $\alpha_n\equiv\alpha+\min\{n-1,M\}$. \footnote{Zamparo et al. (2013) also show that $\{Y_t\}_{t=1}^{\infty}$ is a strictly stationary process. For a general overview of covariance and strict stationarity in GARCH processes, see Francq and Zakoian (2010).}. Below, we will assume the previous form for $\rho\left(\sigma\right)$, since a number of analytical results can be obtained in closed form, see Zamparo et al. (2013). 
For any memory order $M$, the endogenous component $Y_t$ is thus
specified by two parameters only: $\alpha$, determining the form of the
distributions; 
 $\beta$
identifying the scale of the process. The parsimonious number of
parameters, independent of the memory order, is a further advantage
coming from the implementation of the scaling symmetry.

The second component describing the model is partly {\it exogenous} and partly {\it endogenous}
and is introduced as a modulation of the process $Y_t$:
\begin{equation}
X_t=a_{I_t} Y_t
\label{components}
\end{equation}
where the component $a_{I_t}$ is defined as
\begin{equation}
a_{I_t}\equiv\sqrt{I_t^{\;2D}-(I_t-1)^{2D}},
\label{eq_a_i_2}
\end{equation}
with $D$ a suitable parameter.
The process $\{I_t\}_{t=1}^{\infty}$ is a Markov chain in
$\mathbb N^+$, defined by
\begin{equation}
  \pi(i)\equiv\mathbb P[I_1=i]\equiv\nu(1-\nu)^{i-1}
  \label{eqi}
\end{equation}
and
\begin{equation}
  W(i,j)\equiv\mathbb P[I_{t+1}=i|I_t=j]\equiv\nu\delta_{i~1}+(1-\nu)\delta_{i~j+1},
  \label{Wi}
\end{equation}
$\delta_{i~j}$ being the Kronecker delta. Thus, the definition of this
latter component involves two additional parameters:
$0<\nu\leq1$ and $D>0$.  The quantity $\nu^{-1}$ is the average
interval between two random external inputs, where each input event
restarts from $1$ the sequence of integers $I_t$, thus re-setting
$a_{I_t}$ to $1$ (for this reason we will call ``restarts'' these inputs
which could be either endogenous or exogenous).
The exponent $D$ in (\ref{eq_a_i_2}) regulates
how fast the restart is absorbed by market dynamics.
In particular, $0<D<1/2$ corresponds to the case in which the restart
produces a volatility burst due to some external input that then 
decays in time. The component $a_{I_t}$ determines the generalized 
Hurst exponent $H(q)$ associated with the model, which is influenced by both $D$ and $\nu$,
and conveys realistic time properties to the volatility
autocorrelation function.
Since the process $a_{I_t}$ multiplies $Y_t$, it can be regarded as
a Markov process switching the value of the volatility.  In such a
way, the combination of the two components
becomes a Markov-SWitching ARCH (SWARCH) process. Differently from the
proposal of Hamilton and Susmel (1994), the process $I_t$ assumes
values in $\mathbb N^+$ and implies that the switching process is
endowed with an infinite number of states.

One can prove that the composed process $X_t=a_{I_t} Y_t$ is stationary,
ergodic, and displays realistic multiscaling and volatility
autocorrelations if properly calibrated on reasonably long daily returns
series (Zamparo et al. 2013). The estimation of model parameters is performed by matching various
moments of the model's distributions. The model parameters are $5$ in total, $(D,\nu,\alpha,\beta,M)$.
Since $M$ represents a lower
limit for the range up to which memory effects are supposed to be properly 
taken into account by the model, one can fix this parameter in
relation with the time scale relevant for the application of interest. For instance,
in the present work we consider $M=21$, corresponding to a month of market 
activity, and coherently with the common practice adopted for modeling realized volatility sequences, see Corsi (2009). 

\section{Pricing with non-Gaussian scaling}
\label{sec_pricing}
One of the basic features of the BS model (Black and
Scholes, 1973; Merton, 1973; see also Hull, 2000, and references therein) 
is that it derives a closed formula for the
value of, say, a European call option $C$ depending only on two free
parameters: the risk-free interest rate $r$ and the volatility of the
underlying asset $\sigma_{BS}$. 
In principle, these parameters can be identified {\it a priori}
on the basis of historical data.
However, in view of the unrealistic assumption of 
memoryless Gaussian returns, financial practitioners are forced to correct
the BS option value at different strike prices with
appropriately chosen different values of $\sigma_{BS}$. This allows
compensating for volatility changes and for the so-called
``smile effect'' observed in the implicit volatility of the
real-market option value.
From a logical perspective, however,
a model for determining ``derivative prices'' should not rely on 
``market derivative prices'', the latter being what determines the 
implicit volatility.
Instead, derivative prices should be identified by 
a proper rational price (obtained, e.g., through an equivalent
martingale measure) and an efficient hedging strategy
associated with the stochastic dynamics of the
underlying asset (i.e., with the ``physical probability'').

Many models have been proposed to overcome the
BS limitations. Among
others, we cite Heston (1993), Heston and Nandi (2000), Borland and
Bouchaud (2004) and Christoffersen et al. (2006).
On the one hand, approaches like those using continuous-time 
stochastic-volatility models are capable of defining {\it a priori}
formulae for option prices and associated hedging strategies
that reproduce the volatility smile. 
However, they
still need to be calibrated on some real-market derivative prices in order to 
determine the price of other ones. 
On the other hand, pricing methods relying entirely on the underlying 
asset's dynamics typically do not provide explicit formulae. As a consequence,
they require the use of Monte Carlo approaches
that simulate many possible asset's trajectories
to identify efficient hedging strategies and the derivative prices
(see Bouchaud and Potters, 2003).

In view of the fact that the model for the underlying log-returns
presented in the previous section generalizes Gaussian dynamics, 
we wish here to follow a different
approach, which aims at deriving closed expressions for derivative
prices and associated hedging strategies generalizing those of BS to this more complex situation. 
If, consistently with the spirit of our model,  
one assumes that the dynamical evolution of the primary asset's returns could
be described as a mixture of Gaussian processes, a natural way of pricing 
an option and finding a hedging strategy is by averaging the BS formula
through the Gaussian mixture density (see, e.g., Peirano and Challet, 2012).
Our basic idea is then to find the Gaussian mixture density that 
most effectively reproduces the underlying asset's dynamics, given the historical 
conditions observed at the moment the derivative contract is written.
Additional effort must then be made in order to properly take into account 
time restarts that may occur either before the contract stipulation 
or between the writing and the maturity of the contract.
Through this approach 
we will prove that, indeed, we are defining an equivalent martingale price. 
In section 4 we will then show that for S\&P500 options, prices
calculated on the basis of the anomalous scaling dynamics are closer
to market prices than BS ones by about 30\% on average (see
table 4 in section 4). 
We stress that such an evaluation is performed by comparing models
performances on a common playground, without thus correcting the BS
volatility across option maturity and/or strike prices. Adjustments on
the BS inputs would favour the BS model being closer to
the observed prices, thus producing a biased evaluation of the impact of
non-Gaussian scaling on the option prices. 
An assessment of the efficacy of the hedging
strategy is instead left to future work.

\subsection{Closed-form option pricing formulae}
We first point out that, in accordance with the model outlined in Section \ref{sec_model} for the
underlying asset dynamics, our derivation of option prices assumes
discrete time. The market includes a
riskless bond $B_t$ evolving at rate $r>0$ according to
$
B_t=(1+r)^t \; ,
$
and a stock $S_t$ described through its returns $X_t$ and mean
return rate $\mu$, such that
$S_t=S_0 \exp \left[ \mu \, t + \sum_{j=1}^t X_j \right]$, where $S_0$ is the asset value at time $t=0$. We denote
by $\mathbb P$ the ``physical probability measure'' associated with
the returns $X_t$, as described in section \ref{sec_model}, and by
$\mathcal{F}_t$ the $\sigma$-algebra generated by $X_1,\ldots,X_t$,
with $\mathcal{F}_0\equiv\lbrace\emptyset,\Omega\rbrace$.  
$\lbrace S_t\rbrace_{t=0}^T$ is adapted to the filtration $\lbrace
\mathcal{F}_t\rbrace_{t=0}^T$.  
The market is assumed free of arbitrage, liquid, with no
transaction costs, no interest spread, no dividends, and with
unlimited short selling for an unlimited period of time. 

Notice that by choosing
$\rho(\sigma)=\delta(\sigma-\sigma_{BS})$ and $a_{I_t}=1$ for all
$I_t$ (e.g., with $D=1/2$), our model degenerates into a discrete-time
version of the geometric Brownian motion for $S_t$, without memory
effects or volatility bursts. Our basic pricing strategy is
then to define a general equivalent martingale measure which
reproduces (in discrete time) the BS formula in the above
degenerate conditions. We aim first at obtaining equivalent martingale measures
for general (arbitrary) choices of the parameters $\sigma$ and
$a_{i_t}$, that will thus be valid even if they become
random variables. 
In the next subsections, we will identify the most appropriate
distributions for $\sigma$ and $i_t$ conditioned to the past 
dynamics of $X_t$.

Let $f_{t_0,t_1}^I$ be the joint density of $I_t$ in the range
$\left[t_0,t_1\right]$ and $\overline{\rho}$ the volatility density
conditional to some past values for $X_t$ and $I_t$. The following Lemma introduces the martingale measures ${\mathbb P}^\star$ equivalent to $\mathbb P$ used later for option pricing.

\noindent{\bf Lemma 1.} 
{\it 
  Consider the function
  \begin{equation}
    h(x,\sigma,i_t) \equiv \dfrac{1}{\sqrt{2\pi}\;\sigma\;a_{i_t}} 
    \exp \left[ -\frac{1}{2} \left( \frac{x}{\sigma a_{i_t}} + \frac{\sigma a_{i_t}}{2}  
      - \frac{\gamma}{\sigma a_{i_t}} \right)^2 \right] \, ,
  \end{equation}
  with $\gamma \equiv \ln (1+r) - \mu$ and
  $a_{i_t}\equiv\sqrt{i_t^{2D}-(i_t-1)^{2D}}$.
  Given $\sigma\in\mathbb R^+$, $i_t\in\mathbb N^+$ for
  $t=t_0,t_0+1,\ldots,T$ the probability density function\footnote{
  Since we need now to explicitly indicate the time partition to which
  a probability density function is referred to, we slightly change
  the notation with respect to Section 2. 
  In the case of a conditional probability density function, we will explicitly indicate the conditioning values among the arguments after the usual vertical bar, $|$.}
  \begin{equation}
    f_{t_0,T}^{\star\,X}(x_{t_0} , \ldots , x_T|\sigma;i_{t_0},\ldots,i_T)
    = \prod_{t=t_0}^T h(x_t,\sigma,i_t)
  \end{equation}
  defines an equivalent martingale measure ${\mathbb
    P}_{\sigma;i_{t_0},\ldots,i_T}^\star$.\\
  More generally, for any
  given parameters distributions $\overline{\rho}$ and
  $f_{t_0-M,T}^I$, also the probability density function
  \begin{eqnarray}
    &f_{t_0,T}^{\star\,X}(x_{t_0} , \ldots , x_T|\overline{\rho};f_{t_0-M,T}^I) 
    = 
    \displaystyle
    \sum_{i_{t_0-M}=1}^{\infty} \ldots \sum_{i_T=1}^{\infty}\bigg[
      f_{t_0-M,T}^I(i_{t_0-M},\ldots ,i_T)
      \;\cdot &\nonumber\\
      & 
      \displaystyle
      \cdot \;\int_0^{\infty}  
      \left(\prod_{t=t_0}^T h(x_t,\sigma,i_t)\right) 
      \overline{\rho}(\sigma|x_{t_0-M},\ldots,x_{t_0-1};i_{t_0-M},\ldots ,i_T) 
      d\sigma\bigg]&
  \end{eqnarray}
  defines an equivalent martingale measure ${\mathbb
    P}_{\overline{\rho};f_{t_0-M,T}^I}^\star$.
}

The price
$C(K,t_0,T)$ at time $0<t_0<T$ of a European call option with maturity
at time $T$ and strike price $K$ satisfies
$
\dfrac{C(K,t_0,T)}{(1+r)^{t_0}} =  \mathbb{E}_{{\mathbb P}^\star}
\left[ \left.\dfrac{C}{(1+r)^T} \, \right| \, \mathcal{F}_{t_0-1}  \right]  \; ,
$
where $C=C(K,T,T)$ is the option pay-off at maturity
(see, e.g., F\"{o}llmer et al., 2011).
The conditional expectation is taken with respect to the
$\sigma$-algebra $\mathcal{F}_{t_0-1}$ since we assume that while
computing the price $C(K,t_0,T)$, valid at time $t_0$, we still do
not know the value of the return $X_{t_0}$. The following theorem provides the option price.

\noindent{\bf Theorem 1.} 
{\it 
  Given the equivalent martingale measure }
  ${\mathbb P}_{\sigma;i_{t_0},\ldots,i_T}^\star$ \textit{defined in Lemma 1, the call 
  price conditioned by the values} $\sigma$ \textit{and} $i_{t_0},\ldots,i_T$ \textit{is} 
  \begin{equation}
    \label{eq_call_1}
    C(K,t_0,T|\sigma;i_{t_0},\ldots,i_T) = 
    (1+r) \left[ S_{t_0-1} \, \mathrm{N}(d_+) - K (1+r)^{t_0-T-1} \mathrm{N}(d_-) \right], 
  \end{equation}
  \textit{where} $\mathrm{N}(x)$ \textit{is the standard normal cumulative distribution
  function, and}
  \begin{equation}
    d_{\pm} \equiv \dfrac{1}{ \widetilde{\sigma} }\left[ \ln \frac{S_{t_0-1}}{K} 
      + (T-t_0+1)\ln (1+r) \pm \dfrac{(\widetilde{\sigma})^2}{2}\right] \; ,
  \end{equation}
  \begin{equation}
    \label{eq_sigma_prime}
    \widetilde{\sigma} \equiv \widetilde{\sigma} (\sigma ;
    i_{t_0},\ldots,i_T)
    \equiv \sigma \sqrt{a_{i_{t_0}}^2 + \ldots + a_{i_T}^2}\;.
  \end{equation}
  \textit{More generally, given the equivalent martingale measure }
  ${\mathbb P}_{\overline{\rho};f_{t_0-M,T}^I}^\star$\textit{, defined in Lemma 1, the call price
  is}
  \begin{eqnarray}
    &C(K,t_0,T)
    =
    \displaystyle
    \sum_{i_{t_0-M}=1}^{\infty} \ldots \sum_{i_T=1}^{\infty}
    \bigg[
      f_{t_0-M,T}^I(i_{t_0-M},\ldots ,i_T)\;\cdot \qquad\qquad\qquad\qquad&\nonumber\\ 
      \label{eq_call_price}
      &
      \displaystyle
      \cdot \;\int_0^{\infty} 
      \overline{\rho}(\sigma|x_{t_0-M},\ldots,x_{t_0-1};i_{t_0-M},\ldots ,i_T) 
      \;C(K,t_0,T|\sigma;i_{t_0},\ldots,i_T) \; d\sigma \bigg]\;.
  \end{eqnarray}

By assuming $\overline{\rho}(\sigma|x_{t_0-M},\ldots,x_{t_0-1};i_{t_0-M},\ldots ,i_T) 
=\delta(\sigma-\sigma_{BS})$ and $a_{i_t}=1$ for any $i_t$, all
equivalent martingale measures coincide with each other, (\ref{eq_sigma_prime}) simplifies into
$ \widetilde{\sigma} =\sigma_{BS}\;\sqrt{T-t_0+1}$, and the call price obtained in (\ref{eq_call_price})
equals the standard BS formula.
Although not explicitly addressed in the present paper, we highlight that, by means of (\ref{eq_call_1}) and (\ref{eq_call_price}), one also obtains
Delta-hedging formulas which generalize BS' one. 
Namely,
$
\Delta(K,t_0,T|\sigma;i_{t_0},\ldots,i_T) = 
(1+r)\, \mathrm{N}(d_+),
$
and
\begin{eqnarray}
&\Delta(K,t_0,T)
=
\displaystyle
\sum_{i_{t_0-M}=1}^{\infty} \ldots \sum_{i_T=1}^{\infty}
\bigg[
f_{t_0-M,T}^I(i_{t_0-M},\ldots ,i_T)\;\cdot \qquad\qquad\qquad\qquad&\nonumber\\ 
\label{eq_call_delta}
&
\displaystyle
\cdot \;\int_0^{\infty} 
\overline{\rho}(\sigma|x_{t_0-M},\ldots,x_{t_0-1};i_{t_0-M},\ldots ,i_T) 
\;\Delta(K,t_0,T|\sigma;i_{t_0},\ldots,i_T) \; d\sigma \bigg].&
\end{eqnarray}
So far, $f_{t_0-M,T}^I(i_{t_0-M},\ldots ,i_T)$ and
$\overline{\rho}(\sigma|x_{t_0-M},\ldots,x_{t_0-1};i_{t_0-M},\ldots
,i_T)$ are generic distributions for the parameters $i_t$ and
$\sigma$.  In the next subsections, we will link these parameters to
the dynamics characterizing the evolution of $X_t$. This will also
motivate the dependence of the distribution of $\sigma$ on the $x_t$'s
for $t=t_0-M,\ldots,t_{0}-1$ and on the $i_t$'s for
$t=t_0-M,\ldots,T$. Moreover, we will propose for $\overline\rho$ a
Gaussian mixture density representing both the ``state of the market''
summarized in the $M$ returns previous to the derivative contract
pricing day, and the possible future evolutions of the market up to
the maturity time. Since our model contains the hidden component
$I_t$, in order to calculate $\overline\rho$ it is also needed to
include a dependence on the random time $i_t$ both before the contract
pricing day and between the pricing day and the maturity of the
contract.  The density $f_{t_0-M,T}^I$ will then identify the
probability of the random time string $i_{t_0-M},\ldots,i_T$.  Those
two elements are fundamental for the evaluation of
(\ref{eq_call_price}), our main result for derivative pricing. The
evaluation of (\ref{eq_call_price}) will be discussed in a following
subsection (\ref{sec_PriceEval}).

\subsection{Identification of $\overline{\rho}$}
\label{sec_overline_rho}
Within this subsection, the 
sequence $i_{t_0-M},\ldots,i_T$ is assumed to be given. 
At the time $t_0$ at which we wish to price a European option expiring
at time $T$, the ``state of the market'' for the underlying
asset is characterized by the historical
values $x_{t_0-M},\ldots,x_{t_0}$.
Indeed, since $X_t=Y_t\;a_{I_t}$, 
the assumed knowledge of $i_{t_0-M},\ldots,i_{t_0}$ means that we can
identify also $y_{t_0-M},\ldots,y_{t_0}$ as $y_t=x_t/a_{i_t}$.
We aim at finding an approximation for the
probability density function $f_{t_0,T}^X (x_{t_0},\ldots,x_T
\;|\;x_{t_0-1},\ldots,x_{t_0-M};i_{t_0-M},\ldots,i_T)$,
so that the pricing scheme in Section 3.1 could be applied. 
We propose to approximate $f_{t_0,T}^X$ as a Gaussian mixture
$g_{t_0,T}^X$ with mixing density $\overline\rho$:
\begin{eqnarray}
&g_{t_0,T}^X (x_{t_0},\ldots,x_T|x_{t_0-1},\ldots,x_{t_0-M};i_{t_0-M},\ldots,i_T)=\qquad\qquad\qquad\qquad&\nonumber\\
&
\displaystyle
=
\int_0^{\infty} \overline\rho\left(\sigma\;\left|
\;\dfrac{x_{t_0-1}}{a_{i_{t_0-1}}},\ldots,\dfrac{x_{t_0-M}}{a_{i_{t_0-M}}}
;i_{t_0},\ldots,i_{T}\right.\right) 
\;\left[\prod_{t=t_0}^T
\;\dfrac{1}{a_{i_t}}
\;\mathcal{N}_{\sigma} \left(\dfrac{x_t}{a_{i_t}}\right)\right] \; d\sigma \; .&
\end{eqnarray}
In such a way, the returns' dynamic can be viewed as a superposition of Gaussian processes with a fixed
(although stochastic) volatility $\sigma$. Consistently, the mixture
density $\overline\rho$ so identified enters in (\ref{eq_call_price}) to determine the option price. 
The approximation is realized by matching the expected fluctuation of
the return from $t_0$ to $T$ 
calculated through $f_{t_0,T}^X$ with that obtained using 
$g_{t_0,T}^X$. 

\noindent{\bf Theorem 2.} 
\textit{Consider the cumulated return from} $t_0$ \textit{to} $T$, $R\equiv
  X_{t_0}+\cdots+X_T$.
  \textit{If}
  \begin{equation}
    \label{eq_overline_rho_finale}
    \overline\rho\left(\sigma\;\left|
    \;\dfrac{x_{t_0-1}}{a_{i_{t_0-1}}},\ldots,\dfrac{x_{t_0-M}}{a_{i_{t_0-M}}}
    ;i_{t_0},\ldots,i_{T}\right.\right) 
    \equiv 
    \dfrac{
      \displaystyle
      \sum_{t=t_0}^T a_{i_t}^2 
      \;\hat\rho_{t,t}\left(\sigma\;\left|\;\dfrac{x_{t_0-1}}{a_{i_{t_0-1}}},\ldots,
      \dfrac{x_{t_0-M}}{a_{i_{t_0-M}}}\right.\right)
    }
	  {
	    \displaystyle
	    \sum_{t=t_0}^T a^2(i_t)
	  },
  \end{equation}
  \textit{with}
  \begin{equation}
    \tilde{\rho}_{t,t}(\sigma_t\;|\;y_{t-1},\ldots,y_{t-M})\equiv
    \dfrac{\rho(\sigma_t) \prod_{j=t-M}^{t-1} \mathcal{N}_{\sigma_t} (y_j)}
	  {\int_0^{\infty}\;d\sigma_t^\prime\;\rho(\sigma_t^\prime) 
	    \prod_{j^\prime=t-M}^{t-1} \mathcal{N}_{\sigma_t^\prime} (y_{j^\prime})},
	  \label{eq_conditional_rho}
  \end{equation}
  \begin{equation}
    \hat\rho_{t,t}\left(\sigma\left|y_{t_0-1},\ldots,y_{t_0-M}\right.\right)
    =
    \int\prod_{j=t_0}^{t-1} \;\tilde{\rho}_{t,t}\left(\sigma\left|y_{t-1},\ldots,y_{t-M}\right.\right)
    \;f_{j,j}^Y\left(y_j\left|y_{j-1},\ldots,y_{j-M}\right.\right) \; dy_j\;,
  \end{equation}
  \textit{then the expected value of $R^2$ coincides for} $f_{t_0,T}^X$\textit{ and}
  $g_{t_0,T}^X$\textit{:}
  $\mathbb{E}_{f_{t_0,T}^X} \left[ R^2 \right] = \mathbb{E}_{g_{t_0,T}^X} \left[ R^2 \right]$.

\begin{figure}
\includegraphics[width=0.6\columnwidth]{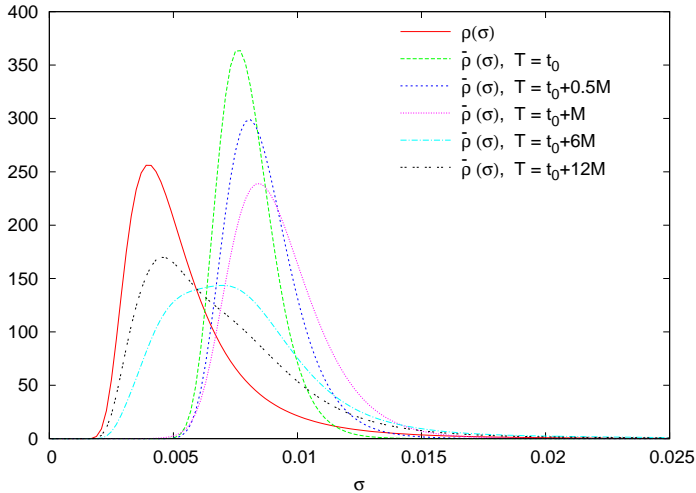}
\caption{Behaviour of $\overline\rho(\sigma)$ at different maturities $T$. $t_0$ is 2 March 2007. Parameters used are $M=21$, $D=1/2$, $\alpha=4$ and $\beta=0.01$. For each $T$, the distribution
$\overline\rho$ has been calculated averaging over $100$
different numerical realizations of the process.}
\label{fig_rhobarra}
\end{figure}

As shown in Fig. \ref{fig_rhobarra}, the
distribution $\overline\rho$ is peaked around the
present value of the volatility if the maturity $T$ is very close to
$t_0$, whereas it becomes
$\overline\rho(\sigma)\simeq\rho(\sigma)$ if $T\gg t_0$.
For
$\rho(\sigma)=\delta(\sigma-\sigma_{BS})$, $\overline\rho$ also
degenerates to a delta,
$\overline\rho(\sigma)=\delta(\sigma-\sigma_{BS})$.  In
practice, $\overline\rho$ can be evaluated by simulating the
process $\lbrace X_t\rbrace_{t=t_0}^T$ conditionally to $M$ past values
$x_{t_0-M},\ldots,x_{t_0-1}$ and given $i_{t_0-M},\ldots,i_T$. 
In the Appendix we give an
explicit expression for $\hat\rho_{t,t}$ when $\rho$ is parametrized as in (\ref{rhoinv}). For the empirical application in 
Section \ref{section_empirical}, for each couple $(t_0,T)$, the distribution
$\overline\rho$ has been evaluated averaging over $100$
different simulated realizations.\footnote{We verified by simulations that $100$ realizations are sufficient to obtain convergence of the distribution.}

\subsection{Identification of $f^I_{t_0-M,T}$}
\label{sec_DerivingfI}
Here we specify a numerical scheme 
which enables the extraction of the random time string
$i_{t_0-M},\ldots,i_T$ and of their associated probability $f^I_{t_0-M,T}$, according to
the model's dynamic and the historical information available at time
$t_0$.
This is instrumental to the evaluation of the option price as discussed in the following subsection. 
Given that the process $I_t$ is not directly identifiable, and given its Markovian nature, we have that the probability of the string $i_{t_0-M},\ldots,i_T$ depends on the whole historical
time series $x_{1},\ldots,x_{t_0-1}$. This prevents the direct evaluation/estimation of the density. To solve this problem we propose a Monte Carlo procedure
which relies on local information only, within a window 
whose size is taken to be $2\tau+1$, to facilitate notations. 
Numerical constraints limit the value of $\tau$ to some units, as specified in what follows. 
More precisely, we present a numerical method to calculate
the probability $f^I_{t_0-M,T}(i_{t_0-M},\ldots,i_T|x_{t_0-M-\tau},\ldots,x_{t_0-1})$.

It is
convenient to distinguish between the distribution
\mbox{$f^I_{t_0-M,t_0-1}$}
of the stochastic
variables $I_{t}$ before the pricing time $t_0$
and the distribution $f^I_{t_0,T}$ of
$I_{t}$ from $t_0$ to $T$. Since the random time
process is a Markov chain, we have
\begin{eqnarray}
\label{eq_f_I_1}
&f^I_{t_0-M,T}(i_{t_0-M},\ldots,i_T|x_{t_0-M-\tau},\ldots,x_{t_0-1}) &\nonumber\\
&= f^I_{t_0-M,t_0-1}(i_{t_0-M},\ldots,i_{t_0-1}|x_{t_0-M-\tau},\ldots,x_{t_0-1}) 
\;f^I_{t_0,T}(i_{t_0},\ldots,i_T|i_{t_0-1}).&
\end{eqnarray}

Before $t_0$, one can exploit the local information contained in the historical returns. In details, we suggest considering 
\begin{eqnarray}\label{eq_f_I_2}
&f^I_{t_0-M,t_0-1}(i_{t_0-M},\ldots,i_{t_0-1}|x_{t_0-M-\tau},\ldots,x_{t_0-1}) 
\simeq\qquad\qquad\qquad\qquad\qquad& \nonumber\\
&\simeq f^I_{t_0-M,t_0-M}(i_{t_0-M}|x_{t_0-M-\tau},\ldots,x_{t_0-M+\tau}) \, \cdot
\qquad\qquad\qquad\qquad\qquad&\nonumber \\
&\displaystyle
\cdot \prod_{t=t_0-M+1}^{t_0-1} f^I_{t,t}\left(
i_t|i_{\max\lbrace t-\tau,t_0-M \rbrace},\ldots,
i_{t-1};x_{t-\tau},\ldots,x_{\min\lbrace t+\tau,t_0-1 \rbrace}\right) \; ,&
\end{eqnarray}
where the approximation comes from to the fact that on the r.h.s. of the equation only local information is considered. 
The appendix shows explicit expressions for 
the r.h.s. of (\ref{eq_f_I_2}); using these
expressions, a Monte Carlo procedure can extract the string 
$i_{t_0-M},\ldots,i_{t_0-1}$
according to our approximation of the probability $f^I_{t_0-M,t_0-1}$.

From $t_0$ to $T$, according to (\ref{Wi}) we simply have
\begin{equation} \label{eq_f_I_3}
f^I_{t_0,T}(i_{t_0},\ldots,i_T|i_{t_0-1}) = \prod_{t=t_0}^{T} W(i_{t},i_{t-1}) \; .
\end{equation}

\subsection{The evaluation of the option price}
\label{sec_PriceEval}
Since we have now specified $\overline\rho$ in (\ref{eq_overline_rho_finale}), and $f_{t_0-M,T}^I$ in (\ref{eq_f_I_1}), (\ref{eq_f_I_2}), and (\ref{eq_f_I_3}), the
option price can be computed according to (\ref{eq_call_price}), which however contains
multiple series with infinite terms.  To evaluate these series
with
respect to the past of $t_0$, we propose the use of a Monte Carlo
method. 
We suggest calculating the option price as the
weighted average on a number of histories $i_{t_0-M},\ldots,i_{t_0-1}$
extracted with probability $f^I_{t_0-M,t_0-1}$. In addition, to reduce the computational complexity, in the range from $t_0$ to $T$, we restrict the sums in (\ref{eq_call_price}) on all
possible realizations $i_{t_0},\ldots,i_T$ with at most two restarts
occurring in the interval $[t_0,T]$. Despite the apparent strong
assumption, we highlight that this corresponds to a second order
approximation of the option price with respect to $\nu$, which is
indeed a small 
parameter. In addition, unreported tests with a third order approximation in $\nu$ (at most three restarts
in $[t_0,T]$) provided negligible changes.

Numerical tests suggest that a local window with $\tau=3$ in (\ref{eq_f_I_2}) is enough to
implement an appropriate Monte Carlo sampling. In the empirical
application outlined in Section 4, the Monte Carlo scheme
has been realized by sampling $N_{MC}$ Monte Carlo realizations, with
$N_{MC}\simeq20$ since $f^I_{t_0-M,t_0-1}$ is sharply
peaked.\footnote{Higher values of $N_{MC}$ do not sensibly affect the option prices.}
Given a Monte Carlo realization $\lbrace
i_{t_0-M}^{(n)},\ldots,i_{t_0-1}^{(n)} \rbrace$ with
$n=1,\ldots,N_{MC},$ and given the random time sequence
$i_{t_0},\ldots,i_{T}$, we can indicate the associated option price
$C^{(n)}$ as
\begin{eqnarray}
&C^{(n)}(K,t_0,T|i_{t_0-M}^{(n)},\ldots,i_{T}^{(n)};i_{t_0},\ldots,i_{T}) 
= \qquad \qquad \qquad \qquad\qquad\qquad &\nonumber \\
&\displaystyle
= \int_0^{\infty} \overline\rho\left(\sigma\;\left|
\;\dfrac{x_{t_0-1}}{a_{i_{t_0-1}^{(n)}}},\ldots,\dfrac{x_{t_0-M}}{a_{i_{t_0-M}^{(n)}}}
;i_{t_0},\ldots,i_{T}\right.\right) \; C(K,t_0,T|\sigma;i_{t_0},\ldots,i_{T}) \; d\sigma \; .&
\label{eq_calculation_option_price_1}
\end{eqnarray}
Therefore, we denote with $C^{(n,0)}$ the option price in the
above expression when no restarts occur between $t_0$ and $T$, while
$C^{(n,1,j)}$ is the option price in the above expression when
only a single restart occurs between $t_0$ and $T$ at time
$t=j$. Finally, $C^{(n,2,j,j^\prime)}$ is the option price in
the above expression when only two restarts occur between $t_0$ and
$T$ at time $t=j$ and $t=j^\prime$. Using those three elements, the
final explicit expression for the option price becomes
\begin{eqnarray}
C(K,t_0,T) = \dfrac{1}{A \; N_{MC}} \sum_{n=1}^{N_{MC}} 
\left[ (1-\nu)^{T-t_0+1} \; C^{(n,0)} (K,t_0,T) +  \right. \nonumber \\
\label{eq_calculation_option_price_2}
\left. 
+  \nu (1-\nu)^{T-t_0}  \sum_{j=t_0}^T C_{t_0,T}^{(n,1,j)} (K)  
+  \nu^2 (1-\nu)^{T-t_0-1} \sum_{j=t_0}^{T-1} \sum_{j^\prime=j+1}^T C_{t_0,T}^{(n,2,j,j^\prime)} (K)  \right] \; ,
\end{eqnarray}
where $A$ is the normalization constant
\begin{equation}
A =  (1-\nu)^{T-t_0+1} + \sum_{j=t_0}^T \nu (1-\nu)^{T-t_0} + \sum_{j=t_0}^{T-1} \sum_{j^\prime=j+1}^T  \nu^2 (1-\nu)^{T-t_0-1}.
\end{equation}
In (\ref{eq_calculation_option_price_2}), the probability of the
random time sequence past to $t_0$, $f^I_{t_0-M,t_0-1}$, has been
replaced by a sum over the Monte Carlo realizations, since the
$i_t^{(n)}$'s are extracted according to $f^I_{t_0-M,t_0-1}$, whereas
the probability of the random time sequence beyond $t_0$,
$f^I_{t_0,T}$, has been explicitly calculated at second order in
$\nu$.

Summarizing, the numerical evaluation of the option
price includes the following steps: (i) Given the historical
information, extract the past random time sequence $i_t^{(n)}$'s
through the Monte Carlo algorithm; (ii) Consider then a possible
future random time sequence with zero, one or at most two restarts
between $t_0$ and $T$; (iii) Evaluate $\overline\rho$ as
described in subsection \ref{sec_overline_rho}; (iv) Calculate a
partial option price using (\ref{eq_calculation_option_price_1});
(v) Take the average across sequences described in (i) and (ii) using,
(\ref{eq_calculation_option_price_2}). 
In terms of CPU time-consumption, points (i) and (iii) are
the bottlenecks in the implementation of this algorithm, and are
independent of the strike price $K$.

We close this section, by noting that the stochastic process previously introduced and the associated option pricing approach allow generating asset returns and option prices characterized by a volatility smile effect, as we will show in the next section.

\section{Empirical application}
\label{section_empirical}
\subsection{Options database}
We now compare the pricing approach previously introduced and the
standard BS model with market prices for options on the S\&P500 index as recovered from Datastream. We consider European call options
with maturity between June 2007 and May 2013, and for those contracts
we download the prices available from January 2007 up to May
2012.\footnote{We consider options with maturity in the twelve months
following the reference download date, the 31 May 2012.} The S\&P500
index options have already been used in several studies dealing with
option pricing, see Bakshi et al. (1997), Ait-Sahalia and Lo (1998),
Dumas et al. (1998), Chernov and Ghysels (2000), Heston and Nandi
(2000) and Christoffersen et al. (2006), among others. Similarly to
Christoffersen et al. (2006), we evaluate the pricing performances
both in-sample and out-of-sample.\footnote{The present paper reports
out-of-sample pricing results only. In-sample results are available
from the authors upon request.} We fix the in-sample to the range
starting in January 2007 and ending at December 2010. The remainder
period, January 2011 to May 2012, is used as out-of-sample. The
Appendix reports additional details on the database and the filters
applied to the option prices, following Bakshi et al. (1997), Dumas et
al. (1998), Ait-Sahalia and Lo (1998), Bakshi et al. (2000),
Christoffersen and Jacobs (2004) and Christoffersen et
al. (2006). Table (\ref{table:outsampleC}) reports the number of
option contracts, the average option price, and the average implied
volatility from the BS model, distinguishing on the basis of the
moneyness (computed as the ratio between the equity index level and
the option strike price) and time to maturity.  Excluding the extreme
moneyness classes (below 0.5 and above 2.5), we observe the presence of
a volatility smile effect, with implied volatility differing across strike prices, and characterized by some asymmetry. In fact, call options highly in-the-money have higher implied volatility compared to out-of-the
money call options. 
Notably, the options' implied volatility
is extremely high, in many cases well above 100\%. This is related to
the peculiar features of the time range analysed in the current paper,
which is characterised by very high volatility.\footnote{Similar
evidences are provided for the in-sample period.}

In comparing our approach to the BS model, option prices are filtered from the dividends effects, as in Bakshi et al. (1997), Ait-Sahalia and Lo (1998) and Christoffersen et al. (2006). In addition, the risk-free rate is proxied by a set of interbank rates, at 1, 3, 6 and 12 months, recovered from Thomson Datastream, and matched with the options maturity; see the Appendix for details.

\subsection{Scaling and persistence in the S\&P500 index}

The pricing of call options starts from the estimation of the model
outlined in section 2.  We use a time-series length of five years to determine the model parameters $\left(D,\nu,\alpha,\beta \right)$, and, for computational simplicity, we impose a grid over the model parameters with precision $5\cdot10^{-3}$ for $D$, $1\cdot10^{-4}$ for $\nu$ and $0.5$ for $\alpha$.\footnote{Parameters have been estimated with a minimum-distance-type estimator, see Zamparo et al. (2013), for additional details.}

\begin{figure}
\begin{center}
\includegraphics[scale=0.9]{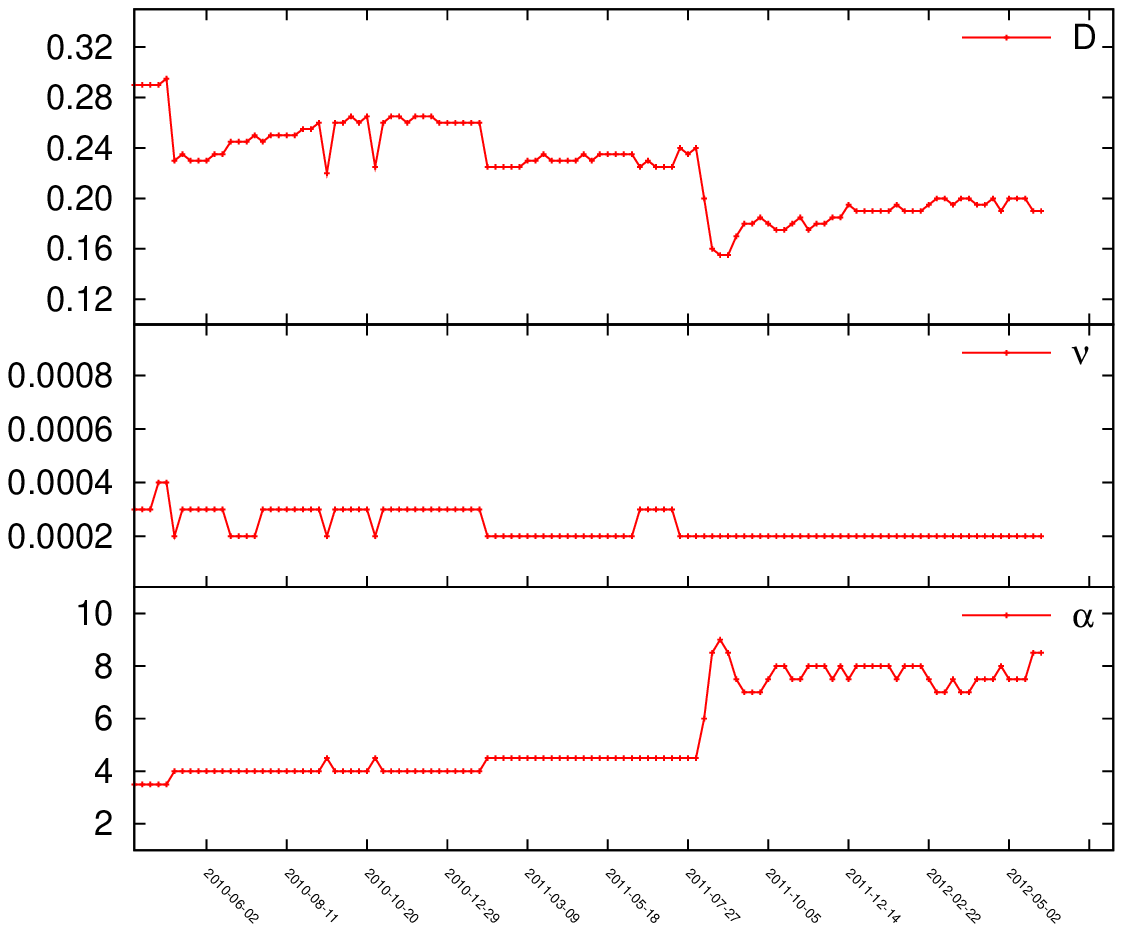} \\
\includegraphics[scale=0.9]{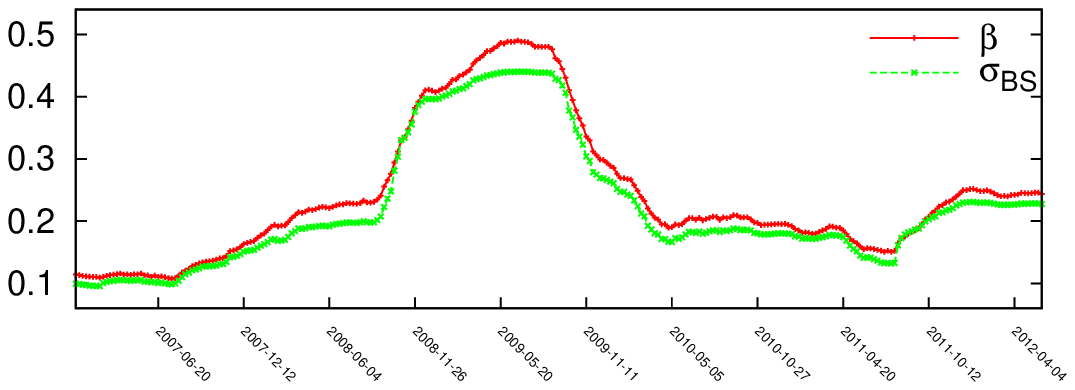}
\caption{Estimated parameters for the out-of-sample analysis; $\beta$ and $\sigma_{BS}$ have been calibrated using the
returns of one year previous to the pricing day, while $D$, $\nu$ and $\alpha$ have been calibrated using the
returns of five years previous to the pricing day.
$D$ (a), $\nu$ (b), $\alpha$ (c), $\beta$ and $\sigma_{BS}$ (d).}
\label{fig_parameters}
\end{center}
\end{figure}

For the in-sample period, the years 2007-2010, we estimate the model
parameters using the S\&P500 index in the range 2006-2010 (five
years). As previously mentioned, we set $M=21$ and the estimated parameters are
$(D,\nu,\alpha)=(0.225,2\cdot10^{-4},4.0)$.
The in-sample parameter $D$ shows evidence of some long-range persistence
in the second order moment of the daily S\&P500 returns. In addition, the shape parameters of the
volatility density are quite high. Finally, the parameter $\nu$
suggests that the average time between two restarts, creating
a regime change, is equal to about 5000 days.\footnote{We also estimated the model with $M=63$ obtaining $(D,\nu,\alpha)=(0.150,8\cdot10^{-4},5.0)$. This suggests an interval between regime changes equal to 1250 days. The parameter $\nu$ changes across values of $M$ and this might be counter-intuitive as it represents a memory-independent part of the model. This is however an estimation/calibration issue. In fact, as in the estimation step we assume {\it a priori} the value of $M$, the estimation of all remaining parameters becomes in fact $M$-dependent. We nevertheless point out that the obtained values for $\nu$ are not much different, and the most relevant outcome is the evidence of having rare regime changes. }
The scale parameter of the volatility density, $\beta$, plays a role
similar to the volatility parameter $\sigma_{BS}$ in the BS model. In
view of a fair comparison with this benchmark model, we estimate these two parameters on the basis of the same sample data. In
preliminary evaluations, we have verified that the use of the above
in-sample interval, 2006-2010, to determine $\sigma_{BS}$ in the BS
formula gives very poor results in comparison with the market option
prices. For this reason, $\beta$ and $\sigma_{BS}$ have been both
estimated in an out-of-sample framework, using the historical data of
one year previous to the pricing day. Using common convention, the
volatility $\sigma_{BS}$ of the BS model is defined as the returns'
standard deviation. The last plot of Figure \ref{fig_parameters}
compares the historical evolution of the $\beta$ (for $M=21$) and of
the BS volatility. We note that the two quantities share a common
pattern, a somewhat expected result given the two parameters are
evaluated over the same sample. In turn, this highlights that the two
models will provide option prices derived from a common
playground. This would have not been the case, for instance, if the BS
prices were derived by changing the underlying volatility according to
the option time to maturity and moneyness. As a consequence, the
differences observed in the option prices reported below are due to
the different dynamic behaviour of the underlying models.

For the out-of-sample evaluation, we estimate the model parameters $D$, $\nu$ and $\alpha$ using the five years data prior to the pricing day.\footnote{The appropriate estimation of those parameters requires longer samples compared to the parameter $\beta$ as discussed in Zamparo et. al (2013).} As a result, we obtain a sequence of parameters values. The parameter evolution is reported in the first to third plots of Figure \ref{fig_parameters}, while some descriptive quantities are included in Table (\ref{table:OutOfSamplePar}) (in all cases we consider $M=21$). We observe that the memory parameter $D$ has a decreasing pattern, suggesting a reduction of the persistence of the equity index. The parameter $\nu$ shows evidence of limited oscillations, influenced by the grid used in its evaluation (its value ranges from $0.0002$ to $0.0004$). Finally, the shape parameter $\alpha$ has a sharp increase in the last part of the sample.

\subsection{Pricing and non-Gaussian scaling}

The pricing approach presented in this paper is contrasted to a more
traditional BS pricing. 

Table (\ref{table:outofsampleCpricing}) reports the out-of-sample
pricing outcomes expressed as root mean squared errors (RMSE) between
the prices implied by the two models and the observed market
prices. The Table shows a preference for the pricing
consistent with the scaling property of the S\&P500. The BS approach
turns out to be preferred for higher levels of moneyness and shorter
maturities. The pricing with anomalous scaling and switching
volatility gives, overall, smaller pricing errors for moneyness levels
up to 1.75, independently of the time to maturity, and for moneyness
levels above 1.75, only for options with maturity in more than three
months. If we focus solely on the moneyness, the introduction of
scaling properties provides pricing errors smaller than those
associated with the BS model for most moneyness levels. Differently,
if we consider only the time to maturity, the BS pricing gives smaller
pricing errors for maturities above three months. Finally, we observe
that the largest differences across the two pricing approaches occur
for moneyness levels below 1.5 independently of the time to
maturity. In those cases, the pricing with anomalous scaling provides
smaller RMSE with a sensibly larger frequency.

\begin{figure}
\begin{center}
\includegraphics{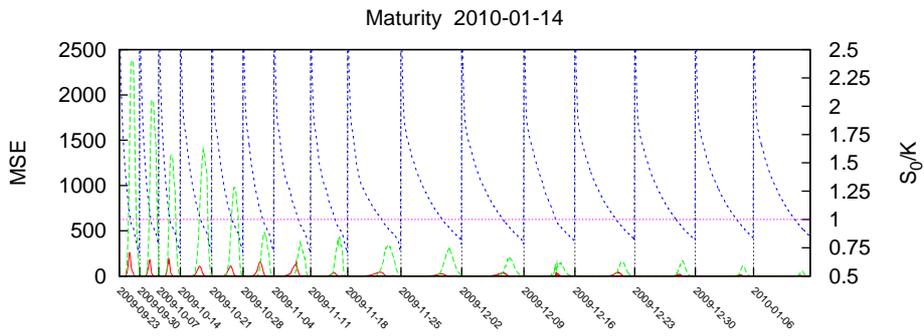}
\end{center}
\caption{Mean squared pricing errors for options with maturity 14 January 2010, across different moneyness levels (blue line - right vertical axis) and pricing days (horizontal axis). Black-Scholes (B\&S - green line - left axis) pricing errors compared to pricing errors from the model with anomalous scaling and switching volatility (ASP - red line - left axis).}\label{fig:pricing_3}
\end{figure}

\begin{figure}
\begin{center}
\subfigure{\includegraphics[scale=0.6]{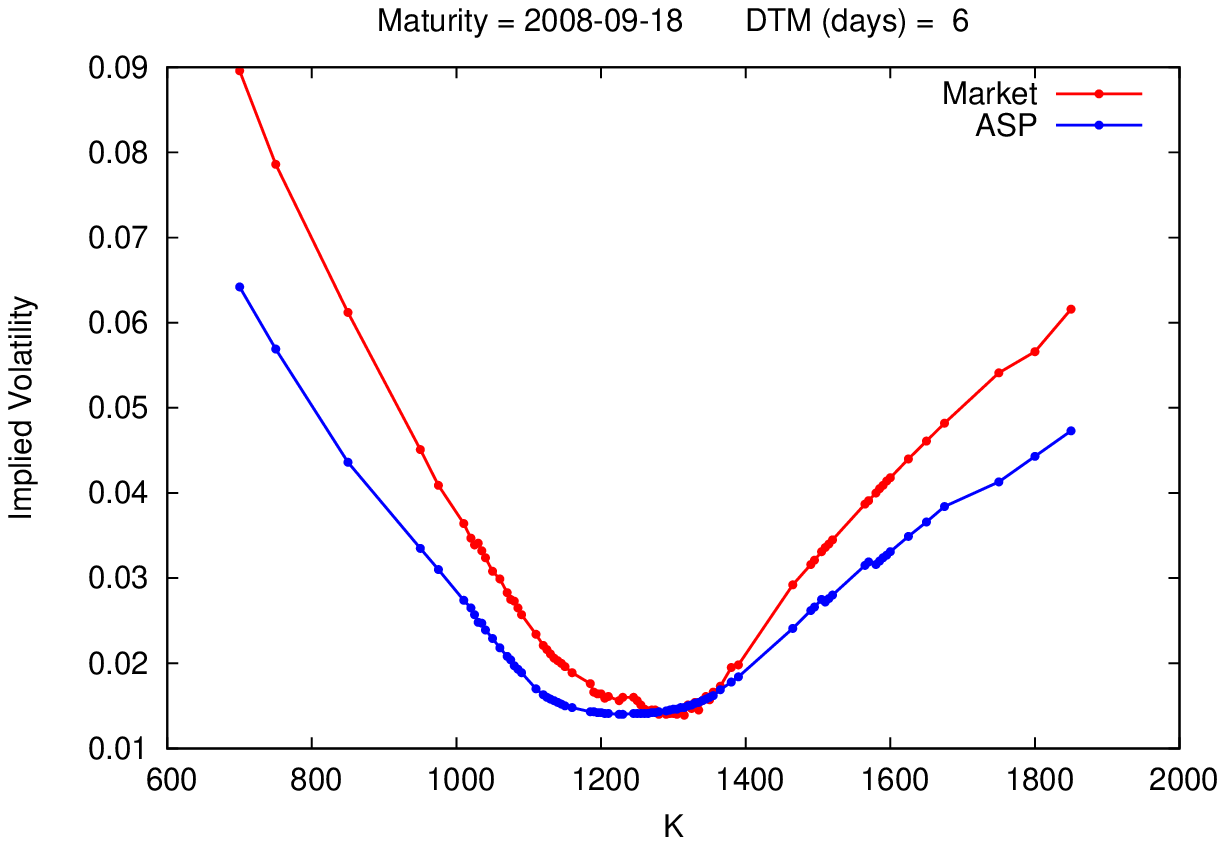}} 
\subfigure{\includegraphics[scale=0.6]{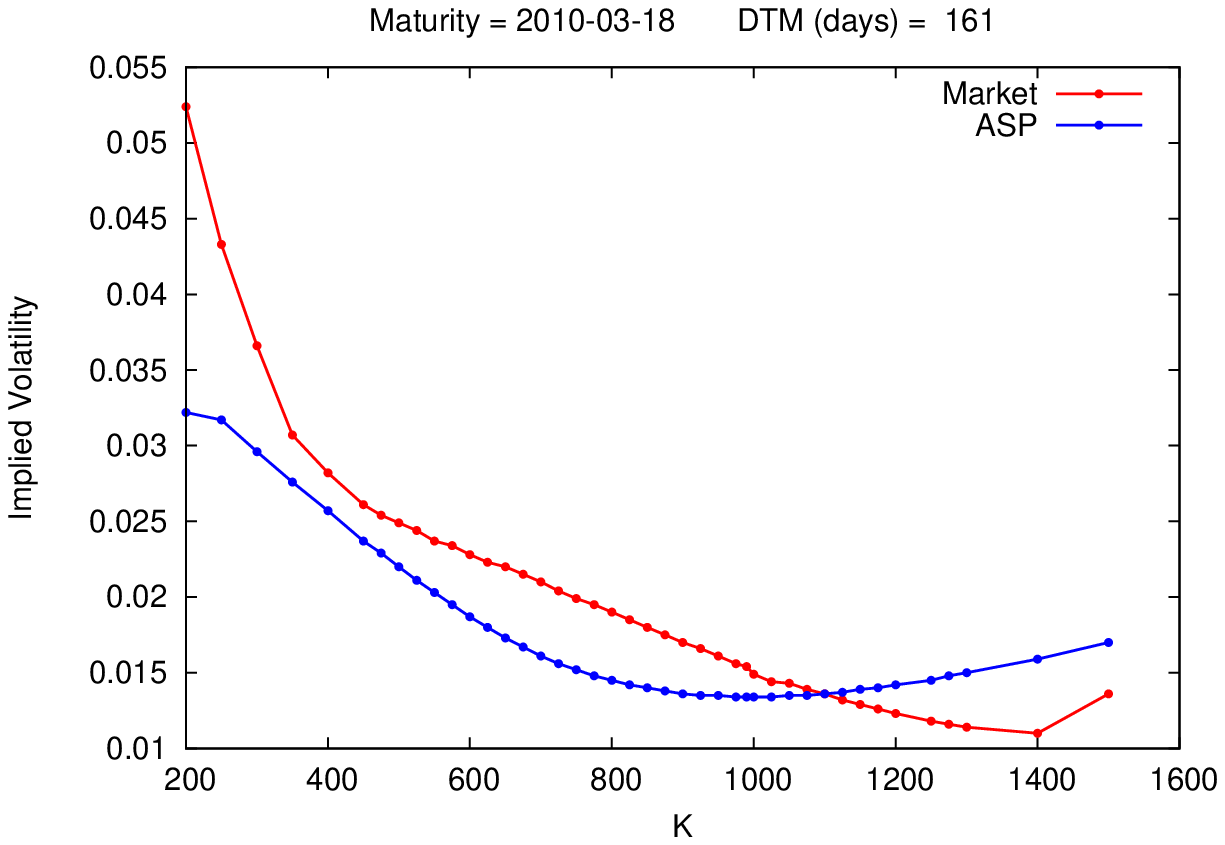}} 
\end{center}
\caption{Implied volatilities associated with option prices determined in presence of anomalous scaling (blue) and empirical smile effect (red) on two different options and days to maturity.}\label{fig:smile_1}
\end{figure}

However, as one might expect, those results are given by a combination
of cases where the preference for our approach is striking, with cases
where the BS pricing is preferred, at least in some dates or for some
maturities. We provide an example in Figures \ref{fig:pricing_3}.\footnote{Additional examples are available upon request.} 
where we report the mean squared error for options
with a given maturity for different moneyness levels and different
pricing days (the Wednesdays of several consecutive weeks). In many cases we observe that the BS model
provides substantially greater pricing errors than those of the anomalous scaling-based prices. Pricing errors might be either increasing or decreasing when approaching time to maturity. 
Finally, the proposed model provides option prices consistent with a volatility smile effect, as shown for a specific cases in Figure \ref{fig:smile_1}.\footnote{Additional examples are available upon request.}

\section{Conclusions}
Self-similarity is a remarkable symmetry linking properties
observed at different scales. Whenever a system
is (at least approximately) self-similar, this symmetry
may be exploited as a guiding line for an appropriate modelling of
non-trivial behaviours (e.g., those induced 
by non-Gaussianity and long-range dependence).

We analysed a stochastic model recently introduced in physics 
(Zamparo et
al., 2013) as a tool for modelling  assets' dynamics
on the basis of the anomalous scaling properties observed on
financial returns. Thanks to the scaling symmetry,
presence of exogenous and endogenous effects,
few parameters, and analytical tractability, 
coexist within the model.   
We discussed the model properties, in particular
with respect to an implementation that amounts to 
an infinite-state, Markov switching, auto-regressive
model. We framed the model components in a financial perspective 
and described the estimation of its parameters. 

Building on the model
potential, we worked out novel closed-form pricing 
and associated hedging strategy formulae
for a European call option. We provided details on
the derivation of the pricing equation and of the associated relevant
quantities. \\ Our work includes an empirical comparison of the
proposed pricing approach with real market prices, together
with benchmark comparisons with a more traditional BS pricing.
We focused on European call options written on the
S\&P500 index, with maturity between 2007 and 2013. 
Results exhibit
evidence that the novel derivative prices are closer to the market prices than BS ones.

Assessment of the efficacy of the proposed strategy, together with the
development of specific Greeks for the aforementioned pricing approach,
are left to future research work.

\smallskip

\textbf{Acknowledgements:} The authors are most grateful to the Editors and two anonymous referees for their insightful comments; The authors acknowledge financial support from the \textit{Fondazione Cassa di Risparmio di Padova e Rovigo} within the \textit{Progetti di Eccellenza 2008-2009} program, project \textit{Anomalous scaling in physics and finance}.

\singlespacing

\begin{table}
\caption{Out-of-sample database description over different levels of moneyness (on rows) and days to maturity (DTM, on columns): number of contracts (upper panel); average CALL option price (central panel); average Black-Scholes implied volatility (lower panel).}
{\footnotesize
\begin{tabular}{lrrrrr}
\hline
DTM (days) & $< 21$ & $21-63$ & $63-126$ & $126-252$ & All \\
Moneyness & & & & & \\
\hline
Number of contracts \\
\hline
$0.50$  & $     2$  & $    11$  & $    30$  & $   117$  & $   160$  \\
$0.50 - 0.75$  & $    28$  & $   155$  & $   494$  & $   925$  & $  1602$  \\
$0.75 - 1.00$  & $  2044$  & $  5561$  & $  2603$  & $  2538$  & $ 12746$  \\
$1.00 - 1.25$  & $  3146$  & $  6437$  & $  2094$  & $  1886$  & $ 13563$  \\
$1.25 - 1.50$  & $  2164$  & $  3987$  & $  1139$  & $  1160$  & $  8450$  \\
$1.50 - 1.75$  & $  1133$  & $  1938$  & $   725$  & $   799$  & $  4595$  \\
$1.75 - 2.00$  & $   459$  & $   738$  & $   436$  & $   548$  & $  2181$  \\
$2.00 - 2.25$  & $   144$  & $   319$  & $   291$  & $   386$  & $  1140$  \\
$2.25 - 2.50$  & $    76$  & $   167$  & $   192$  & $   289$  & $   724$  \\
$> 2.50$  & $   396$  & $   876$  & $   898$  & $  1326$  & $  3496$  \\
All   & $  9592$  & $ 20189$  & $  8902$  & $  9974$  & $ 48657$  \\
\hline
Average option price \\
\hline
$< 0.50$  & $    0.05$  & $    0.09$  & $    0.29$  & $    0.35$  & $    0.32$  \\
$0.50 - 0.75$  & $    0.25$  & $    0.24$  & $    0.41$  & $    0.65$  & $    0.53$  \\
$0.75 - 1.00$  & $    4.36$  & $    9.30$  & $   14.81$  & $   28.20$  & $   13.39$  \\
$1.00 - 1.25$  & $  133.46$  & $  138.46$  & $  144.14$  & $  172.24$  & $  142.87$  \\
$1.25 - 1.50$  & $  342.31$  & $  339.55$  & $  335.12$  & $  342.80$  & $  340.11$  \\
$1.50 - 1.75$  & $  487.86$  & $  481.70$  & $  478.94$  & $  478.28$  & $  482.19$  \\
$1.75 - 2.00$  & $  603.35$  & $  598.90$  & $  586.95$  & $  578.33$  & $  592.28$  \\
$2.00 - 2.25$  & $  685.49$  & $  682.29$  & $  672.24$  & $  654.81$  & $  670.82$  \\
$2.25 - 2.50$  & $  748.82$  & $  744.33$  & $  731.36$  & $  707.02$  & $  726.47$  \\
$> 2.50$  & $  962.03$  & $  958.83$  & $  929.54$  & $  899.65$  & $  929.22$  \\
All   & $  264.37$  & $  240.44$  & $  280.41$  & $  315.20$  & $  267.79$  \\
\hline
Average implied volatility \\
\hline
$< 0.50$ &  & 0.581 & 0.447 & 0.337 & 0.380 \\
$0.50 - 0.75$ & 0.675 & 0.317 & 0.256 & 0.188 & 0.227 \\
$0.75 - 1.00$ & 0.186 & 0.173 & 0.171 & 0.177 & 0.175 \\
$1.00 - 1.25$ & 0.300 & 0.257 & 0.247 & 0.246 & 0.265 \\
$1.25 - 1.50$ & 0.498 & 0.353 & 0.313 & 0.301 & 0.383 \\
$1.50 - 1.75$ & 0.646 & 0.422 & 0.330 & 0.316 & 0.456 \\
$1.75 - 2.00$ & 0.745 & 0.492 & 0.347 & 0.304 & 0.493 \\
$2.00 - 2.25$ & 0.813 & 0.493 & 0.356 & 0.302 & 0.459 \\
$2.25 - 2.50$ & 0.904 & 0.542 & 0.389 & 0.315 & 0.475 \\
$> 2.50$ & 1.430 & 0.790 & 0.508 & 0.329 & 0.650 \\
All & 0.448 & 0.307 & 0.285 & 0.256 & 0.326 \\
\hline
\end{tabular}
}
\label{table:outsampleC}
\end{table}


\begin{table}
\caption{Estimated parameters for the out-of-sample analysis. $\beta$ and $\sigma_{BS}$ have been calibrated using the
returns of one year previous to the pricing day, while $D$, $\nu$ and $\alpha$ have been calibrated using the
returns of five years previous to the pricing day and $M=21$.}
{\footnotesize
\begin{tabular}{lrrrrr}
\hline
 & $D$ & $\nu$ & $\alpha$ & $\beta$ & $\sigma_{BS}$ \\
\hline
mean     & 0.224 & 0.0002 & 5.50 & 0.241 & 0.221 \\
st. dev. & 0.033 & 0.0001 & 1.75 & 0.108 & 0.101 \\
min.     & 0.155 & 0.0002 & 3.50 & 0.108 & 0.096 \\
max.     & 0.295 & 0.0004 & 9.00 & 0.490 & 0.440 \\
median   & 0.230 & 0.0002 & 4.50 & 0.207 & 0.187 \\
\hline
\end{tabular}
}
\label{table:OutOfSamplePar}
\end{table}


\begin{table}
\caption{Out-of-sample pricing root mean squared errors between the option prices derived from the two alternative pricing approaches (the one described in the present paper and the traditional Black-Scholes) and the option prices observed in the market over different levels of moneyness, reported over rows, and different days to maturity (DTM), reported over columns.}
{\footnotesize
\begin{tabular}{lrrrrr}
\hline
DTM (days) & $< 21$ & $21-63$ & $63-126$ & $126-252$ & Total \\
Moneyness & & & & & \\
\hline
\multicolumn{6}{l}{Pricing with anomalous scaling and switching volatility} \\
\hline
$<0.50$  & $    0.0500$  & $    0.1091$  & $    0.3736$  & $    0.4276$  & $    0.4009$  \\
$0.50 - 0.75$  & $    0.4204$  & $    0.5869$  & $    0.8635$  & $    2.7064$  & $    2.1203$  \\
$0.75 - 1.00$  & $    3.9723$  & $    4.5316$  & $    7.1004$  & $   14.5335$  & $    7.9903$  \\
$1.00 - 1.25$  & $    4.5711$  & $    9.2425$  & $   16.3600$  & $   33.7892$  & $   15.6675$  \\
$1.25 - 1.50$  & $    1.8501$  & $    4.0428$  & $    7.8244$  & $   22.3663$  & $    9.2474$  \\
$1.50 - 1.75$  & $    1.4836$  & $    3.2287$  & $    7.6356$  & $   17.2082$  & $    8.1012$  \\
$1.75 - 2.00$  & $    1.5083$  & $    2.9552$  & $    8.2243$  & $   13.9656$  & $    8.1217$  \\
$2.00 - 2.25$  & $    1.5855$  & $    2.7726$  & $    8.4971$  & $   12.8151$  & $    8.7467$  \\
$2.25 - 2.50$  & $    1.6387$  & $    2.7528$  & $    8.9124$  & $   11.5236$  & $    8.7236$  \\
$> 2.50$  & $    1.7663$  & $    2.9877$  & $    4.4880$  & $    6.0581$  & $    4.6566$  \\
All   & $    3.3978$  & $    6.1656$  & $    9.9871$  & $   19.4427$  & $   10.6670$  \\
\hline
\multicolumn{6}{l}{Pricing with Black-Scholes} \\
\hline
$<0.50$  & $    0.0500$  & $    0.1158$  & $    0.3898$  & $    0.4810$  & $    0.4457$  \\
$0.50 - 0.75$  & $    0.4328$  & $    0.3490$  & $    0.5561$  & $    1.9711$  & $    1.5342$  \\
$0.75 - 1.00$  & $    4.8791$  & $    7.9976$  & $    9.0545$  & $   12.4562$  & $    8.9085$  \\
$1.00 - 1.25$  & $    5.5124$  & $   10.0431$  & $   13.9941$  & $   24.8824$  & $   13.0862$  \\
$1.25 - 1.50$  & $    1.8571$  & $    4.3065$  & $    8.4834$  & $   22.1285$  & $    9.3035$  \\
$1.50 - 1.75$  & $    1.4775$  & $    3.2480$  & $    7.7517$  & $   17.6818$  & $    8.2966$  \\
$1.75 - 2.00$  & $    1.5057$  & $    2.9546$  & $    8.2606$  & $   14.1985$  & $    8.2296$  \\
$2.00 - 2.25$  & $    1.5842$  & $    2.7720$  & $    8.5098$  & $   12.9228$  & $    8.8033$  \\
$2.25 - 2.50$  & $    1.6384$  & $    2.7522$  & $    8.9198$  & $   11.5729$  & $    8.7516$  \\
$> 2.50$  & $    1.7661$  & $    2.9875$  & $    4.4889$  & $    6.0673$  & $    4.6614$  \\
Total   & $    4.0462$  & $    7.4394$  & $    9.6756$  & $   16.2858$  & $    9.8836$  \\
\hline
\end{tabular}
}
\label{table:outofsampleCpricing}
\end{table}


\footnotesize

\section*{Appendix}

\subsection*{Proof of Lemma 1.}

For all
the measures ${\mathbb P}^\star$ introduced in Lemma 1 and for $t_0\le t\le T$, it is
    straightforward to see that \\
    \mbox{$
\mathbb{E}_{{\mathbb P}^\star} \left[\left. \frac{S_t}{(1+r)^t} \, \right| \, \mathcal{F}_{t-1}  \right] = \frac{S_{t-1}}{(1+r)^{t-1}}
$}.$\quad\square$

\subsection*{Proof of Theorem 1.}

Using the equivalent martingale measure 
${\mathbb P}_{\sigma;i_{t_0},\ldots,i_T}^\star$, we have
\begin{eqnarray}
C(K,t_0,T|\sigma;i_{t_0},\ldots,i_T) = (1+r)^{t_0-T} \, 
\mathbb{E}_{{\mathbb P}_{\sigma;i_{t_0},\ldots,i_T}^\star} 
       [\max \lbrace S_T -K,0 \rbrace \, | \, \mathcal{F}_{t_0-1}] = \nonumber \\
= (1+r)^{t_0-T} \, \int_{-\infty}^{+\infty}  \max \lbrace S_T -K,0 \rbrace \; 
\left(\prod_{t=t_0}^T h(x_t,\sigma,i_t)\right) \; dx_{t_0} \, \ldots \, dx_T .
\end{eqnarray}
Since $S_T = S_{t_0-1} \exp\left[\mu (T-t_0+1) + \sum_{t=t_0}^T
  X_t\right]$, making the change of variables \mbox{$\lambda_t \equiv
  x_t - \gamma$} we obtain
\begin{eqnarray}
C(K,t_0,T|\sigma;i_{t_0},\ldots,i_T) = 
(1+r) \, \int_{-\infty}^{+\infty}  
\max \left\lbrace  S_{t_0-1} e^{\sum_{t=t_0}^T \lambda_t} - K (1+r)^{t_0-T-1},0 \right\rbrace  \nonumber \\
 \qquad \left( \prod_{t=t_0}^T \dfrac{1}{\sqrt{2\pi} \sigma a_{i_t}} \exp \left[ -\frac{1}{2} \left(  
\frac{\lambda_t}{\sigma a_{i_t}} + \frac{\sigma a_{i_t}}{2} \right)^2 \right] \right) \; d\lambda_{t_0} \, \ldots \, d\lambda_T = \nonumber \\
= (1+r) \, \int_{-\infty}^{+\infty} \max \left\lbrace  S_{t_0-1} e^{\lambda} - K (1+r)^{t_0-T-1} ,0 \right\rbrace 
\dfrac{1}{\sqrt{2\pi} \; \widetilde{\sigma} } \exp \left[ -\frac{1}{2} \left(  
\frac{\lambda}{\widetilde{\sigma}} + \frac{\widetilde{\sigma}}{2} \right)^2 \right] \; d\lambda \, , \;
\end{eqnarray}
where for the last equality we have used the closeness of the Gaussian under aggregation with $\lambda \equiv \sum_{t=t_0}^T
\lambda_t$.
Following now standard steps as in the BS derivation, we
end up with (\ref{eq_call_1}). \\
\noindent Taking the expectation over the distributions of $\sigma$ and
$i_{t_0},\ldots,i_T$, (\ref{eq_call_price}) derives then from (\ref{eq_call_1}) . $\quad\square$

\subsection*{Proof of Theorem 2.}

We have 
\begin{eqnarray}
&f_{t_0,T}^X (x_{t_0},\ldots,x_T \;|\;x_{t_0-1},\ldots,x_{t_0-M};i_{t_0-M},\ldots,i_T) 
=&\nonumber\\
&\qquad\qquad\displaystyle\prod_{t=t_0}^T \; f_{t,t}^X 
(x_t\;|\;x_{t-1},\ldots,x_{t_0-M};i_{t_0-M},\ldots,i_T).& 
\end{eqnarray}
Using $X_t=a_{I_t}\;Y_t$ and conditionally to the sequence $i_{t_0-M},\ldots,i_T$, 
for $t_0\leq t\leq T$ we can write
\begin{eqnarray}
f_{t,t}^X (x_t \;|\;x_{t-1},\ldots,x_{t_0-M};i_{t_0-M},\ldots,i_T)
&=& 
\dfrac{1}{a_{i_t}}\;f_{t,t}^Y\left(\left.\dfrac{x_t}{a_{i_t}} \;\right|
\;\dfrac{x_{t-1}}{a_{i_{t-1}}},\ldots,\dfrac{x_{t_0-M}}{a_{i_{t_0-M}}}\right)
\nonumber\\
&=&
\dfrac{1}{a_{i_t}}\;f_{t,t}^Y\left(\left.\dfrac{x_t}{a_{i_t}} \;\right|
\;\dfrac{x_{t-1}}{a_{i_{t-1}}},\ldots,\dfrac{x_{t-M}}{a_{i_{t-M}}}\right),
\end{eqnarray}
where we have used the fact that $Y_t$
depends on the previous $M$ values only.
Assuming, for simplicity of
notation that $t_0\geq M$, setting $y=x/a$ and exploiting (\ref{gs}) and (\ref{gp_1}) we get
\begin{equation}
\label{eq_end_1}
f_{t,t}^Y (y_t \;|\;y_{t-1},\ldots,y_{t-M})
=
\int_0^{\infty} \;\hat\rho_{t,t}(\sigma_t\;|\;y_{t-1},\ldots,y_{t-M}) \; \mathcal{N}_{\sigma_t} (y_t) \; d\sigma_t ,
\end{equation}
with $\hat\rho_{t,t}(\sigma_t\;|\;y_{t-1},\ldots,y_{t-M})$ given by (\ref{eq_conditional_rho}).
Putting together, we obtain
\begin{eqnarray}
&f_{t_0,T}^X (x_{t_0},\ldots,x_T|x_{t_0-1},\ldots,x_{t_0-M};i_{t_0-M},\ldots,i_T)=\qquad\qquad\qquad\qquad&\nonumber\\
\label{eq_with_rho_hat}
&
\displaystyle
=
\prod_{t=t_0}^T
\;\dfrac{1}{a_{i_t}}
\;\int_0^{\infty} \hat\rho_{t,t}\left(\sigma_t\;\left|\;\dfrac{x_{t-1}}{a_{i_{t-1}}},\ldots,\dfrac{x_{t-M}}{a_{i_{t-M}}}\right.\right) \;\mathcal{N}_{\sigma_t} \left(\dfrac{x_t}{a_{i_t}}\right) \; d\sigma_t .&
\end{eqnarray}
Notice that (\ref{eq_with_rho_hat}) is exact but has the disadvantage that
$\hat\rho_{t,t}$ is a conditional density for a ``dynamical stochastic
volatility'' $\sigma_t$ that also depends on the endogenous values
$y_t=x_t/a_{i_t}$ {\it future} to $t_0$ and is not then identified by
historical data only.
Using the fact that $\frac{1}{a_{i_t}}\mathcal{N}_{\sigma_t} \left( \dfrac{x_t}{a_{i_t}}  \right)  = \mathcal{N}_{a_{i_t} \sigma_t} (x_t)$ and the parity of $f_{t_0,T}^X$ with respect to all its arguments, we have
\begin{eqnarray}
\langle R^2 \rangle_{f_{t_0,T}^X} &=& \int dx_{t_0} \ldots dx_T \; (x_{t_0} + \ldots + x_T)^2 \cdot \nonumber \\
& & \qquad \cdot \; \prod_{t=t_0}^T 
\;\int_0^{\infty}  \hat\rho_{t,t}\left(\sigma_t\;\left|\;\dfrac{x_{t-1}}{a_{i_{t-1}}},\ldots,\dfrac{x_{t-M}}{a_{i_{t-M}}}\right.\right) \;\mathcal{N}_{a_{i_t} \sigma_t} (x_t) \; d\sigma_t \;
 = \nonumber \\
&=& \int dx_{t_0} \ldots dx_T \; (x_{t_0}^2 + \ldots + x_T^2) \; \cdot \nonumber \\
& & \qquad \cdot \; \prod_{t=t_0}^T 
\;\int_0^{\infty} \hat\rho_{t,t}\left(\sigma_t\;\left|\;\dfrac{x_{t-1}}{a_{i_{t-1}}},\ldots,\dfrac{x_{t-M}}{a_{i_{t-M}}}\right.\right) \;\mathcal{N}_{a_{i_t} \sigma_t} (x_t) \; d\sigma_t  \;
 = \nonumber \\
&=& \sum_{t=t_0}^T dx_t \; x_t^2  \; \int_0^{\infty} \hat\rho_{t,t}\left(\sigma_t\;\left|\;\dfrac{x_{t_0-1}}{a_{i_{t_0-1}}},\ldots,\dfrac{x_{t_0-M}}{a_{i_{t_0-M}}}\right.\right) \;\mathcal{N}_{a_{i_t} \sigma_t} (x_t) \; d\sigma_j \;
 = \nonumber \\
&=&  \sum_{t=t_0}^T a^2_{i_t} \langle \sigma^2 \rangle_{\hat\rho_{t,t}}
\end{eqnarray}
where we use the facts that
\begin{eqnarray}
\hat\rho_{t,t} ( \sigma_t\; | \;y_{t_0-1},\ldots,y_{t_0-M} ) = \qquad \qquad \qquad \qquad \qquad \qquad \qquad \qquad \nonumber \\
= \int \hat\rho_{t,t} ( \sigma_t \; | \; y_{t-1},\dots, y_{t-M} ) \;
\left[ \, \prod_{\stackrel{j=t_0}{j\neq t}}^T 
\hat\rho_{j,j} ( \sigma_j\; | \;y_{j-1},\ldots,y_{j-M} ) \; \mathcal{N}_{\sigma_j} (y_j) \; dx_j \right] \; ,
\end{eqnarray}
\begin{equation}
\int_{-\infty}^{\infty} z^2 \mathcal{N}(z) \; dz = 1
\end{equation}
and
\begin{equation}
\langle \sigma^2 \rangle_{\hat\rho_{t,t}} \equiv \int_0^{\infty} \sigma_t^2 \; \hat\rho_{t,t}\left(\sigma_t\;\left|\;\dfrac{x_{t_0-1}}{a_{i_{t_0-1}}},\ldots,\dfrac{x_{t_0-M}}{a_{i_{t_0-M}}}\right.\right) \; d\sigma_t \; .
\end{equation}

On the other hand:
\begin{eqnarray}
\langle R^2 \rangle_{g_{t_0,T}^X} &=& \int dx_{t_0} \ldots dx_T \; (x_{t_0} + \ldots + x_T)^2 \cdot \nonumber \\
& & \quad \int_0^{\infty} 
\overline\rho\left(\sigma\;\left|
\;\dfrac{x_{t_0-1}}{a_{i_{t_0-1}}},\ldots,\dfrac{x_{t_0-M}}{a_{i_{t_0-M}}}
;i_{t_0},\ldots,i_{T}\right.\right) 
\;\left[\prod_{t=t_0}^T
\;\dfrac{1}{a_{i_t}}
\;\mathcal{N}_{\sigma} \left(\dfrac{x_t}{a_{i_t}}\right)\right] \; d\sigma \;= \nonumber \\
&=& \int_0^{\infty} 
\overline\rho\left(\sigma\;\left|
\;\dfrac{x_{t_0-1}}{a_{i_{t_0-1}}},\ldots,\dfrac{x_{t_0-M}}{a_{i_{t_0-M}}}
;i_{t_0},\ldots,i_{T}\right.\right) \;
\left[ \sum_{t=t_0}^T \int  x_t^2 \; \mathcal{N}_{a_{i_t} \sigma} (x_t) \; dx_t  \right]  d\sigma \; = \nonumber \\
&=& \langle \sigma^2 \rangle_{\overline\rho} \; \sum_{t=t_0}^T a^2_{i_t} \; ,
\end{eqnarray}
where
\begin{equation}
\langle \sigma^2 \rangle_{\overline\rho} \equiv \int_0^{\infty}  \sigma^2 \; \overline\rho\left(\sigma\;\left|
\;\dfrac{x_{t_0-1}}{a_{i_{t_0-1}}},\ldots,\dfrac{x_{t_0-M}}{a_{i_{t_0-M}}}
;i_{t_0},\ldots,i_{T}\right.\right) \; d\sigma \; .
\end{equation}
By direct inspection, we thus see that 
$\langle R^2 \rangle_{f_{t_0,T}^X} = \langle R^2 \rangle_{g_{t_0,T}^X}$.
$\quad\square$

\subsection*{An explicit expression for $\hat{\rho}_{t,t}$}
By choosing
the density $\rho(\sigma)$ to be an inverse-Gamma distribution, (\ref{rhoinv}), it is possible to analytically solve (\ref{gp_1}) and
(\ref{eq_conditional_rho}). 
Indeed, with an inverse-Gamma distribution for $\rho(\sigma)$, (\ref{gp_1}) becomes
\begin{equation}
\varphi_t(y_1,\ldots,y_t) = \dfrac{\beta^{\alpha} \, \Gamma\left( \frac{\alpha + t}{2}\right) }{\pi^{\frac{t}{2}} \; \Gamma\left( \frac{\alpha}{2}\right)} \left[ \beta^2 + y_1^2 + \ldots + y_t^2 \right]^{-\frac{\alpha + t}{2}} \; ,
\end{equation}
and (\ref{eq_end_1}) casts into
\begin{equation}
f_{t,t}^Y (y_t \;|\;y_{t-1},\ldots,y_{t-M})
=\dfrac{\Gamma \left( \frac{\alpha + M + 1}{2} \right) }{ \sqrt{\pi} \; \Gamma \left( \frac{\alpha + M}{2} \right)} \; \dfrac{\left[ \beta^2 + y_{t-M}^2 + \ldots + y_t^2 \right] ^{-\frac{\alpha + M + 1}{2}}}   {\left[ \beta^2 + y_{t-M}^2 + \ldots + y_{t-1}^2 \right] ^{-\frac{\alpha + M}{2}}}
\end{equation}
and finally (\ref{eq_conditional_rho}) now reads
\begin{equation}
\hat\rho_{t,t}(\sigma_t\;|\;y_{t-1},\ldots,y_{t-M}) = 
\dfrac{2 \; e^{-\frac{\beta^2 + y_{t-M}^2 + \ldots + y_{t-1}^2}{2\sigma_t^2}} }{2^{\frac{\alpha+M}{2}} \; \sigma_t^{\alpha +M+1} \; \Gamma \left( \frac{\alpha + M}{2} \right)}
\; \left[ \beta^2 + y_{t-M}^2 + \ldots + y_{t-1}^2 \right] ^{\frac{\alpha + M}{2}} \, .
\end{equation}

\subsection*{An explicit expression for $f_{t_0-M,t_0-1}^I$}
In order to write explicitly the r.h.s. of (\ref{eq_f_I_2}), 
let us start with the model's probability density function
of $t$ return variables $X_1,\ldots,X_{t}$ for $t \le M+1$ (see
Zamparo et al., 2013, for details):
\begin{equation}
f^X_{1,t} (x_1,\ldots,x_{t}) = \sum_{i_1=1}^{\infty} \ldots \sum_{i_{t}=1}^{\infty} f^{X,I}_{1,t} (i_1,\ldots,i_{t};x_1,\ldots,x_{t}) \; ,
\end{equation}
where
\begin{equation}
 f^{X,I}_{1,t}(i_1,\ldots,i_{t};x_1,\ldots,x_{t}) \equiv  \dfrac{W(i_{t},i_{t-1}) \ldots W(i_2,i_1) \pi(i_1)}{a_{i_1} \ldots a_{i_{t}}} \;
\varphi_{t} \left( \dfrac{x_1}{a_{i_1}} , \ldots , \dfrac{x_{t}}{a_{i_{t}}} \right) \, .
\end{equation}
With a slightly more complex notation, let us also define:
\begin{equation}
f^{X,I}_{1,t \, ; \, t_1,t_2} (i_{t_1},\ldots ,i_{t_2};x_1,\ldots,x_{t}) \equiv \sum_{i_1=1}^{\infty} \ldots \sum_{i_{t_1-1}=1}^{\infty} \; \; \sum_{i_{t_2+1}=1}^{\infty} \ldots \sum_{i_{t}=1}^{\infty} \; f^{X,I}_{1,t} (i_1,\ldots,i_{t};x_1,\ldots,x_{t}) \; ,
\end{equation}
where $1 \le t_1 \le t_2 \le t$.

We can thus specify the factors in the r.h.s. of (\ref{eq_f_I_2}) as
\begin{eqnarray}
f^I_{t_0-M,t_0-M}(i_{t_0-M}|x_{t_0-M-\tau},\ldots,x_{t_0-M+\tau}) = \qquad \qquad \qquad \nonumber \\
= \dfrac{f^{X,I}_{t_0-M-\tau,t_0-M+\tau \, ; \, t_0-M, t_0-M}(i_{t_0-M};x_{t_0-M-\tau},\ldots,x_{t_0-M+\tau})}{f^X_{t_0-M-\tau,t_0-M+\tau}(x_{t_0-M-\tau},\ldots,x_{t_0-M+\tau})} 
\end{eqnarray}
and
\begin{eqnarray}
f^I_{t,t}\left(
i_t|i_{t-1};x_{t-\tau},\ldots,x_{\min\lbrace t+\tau,t_0-1 \rbrace}\right) = \qquad \qquad \qquad \nonumber \\
= \dfrac{f^{X,I}_{t-\tau,\min \lbrace t+\tau,t_0+1 \rbrace \, ; \, \max\lbrace t-\tau,t_0-M \rbrace , t}(i_{\max\lbrace t-\tau,t_0-M \rbrace},\ldots, i_{t};x_{t-\tau},\ldots,x_{\min\lbrace t+\tau,t_0-1 \rbrace})}
{f^{X,I}_{t-\tau,\min \lbrace t+\tau,t_0+1 \rbrace \, ; \, \max\lbrace t-\tau,t_0-M \rbrace , t-1}(i_{\max\lbrace t-\tau,t_0-M \rbrace},\ldots, i_{t-1};x_{t-\tau},\ldots,x_{\min\lbrace t+\tau,t_0-1 \rbrace})} \, . \nonumber \\
\end{eqnarray}

\subsection*{Option database filters and data management}
Similarly to Dumas et al. (1998), Christoffersen and Jacobs (2004), and Christoffersen et al. (2006), we focus on the Wednesday prices to reduce the computational burden associated with the pricing of several options with different maturities and strike prices. We then apply to the raw Wednesday data filters similar to those proposed by Bakshi et al. (1997): we discard the last week of trading for each option contract to limit the effects associated with option expiration; we drop prices below a threshold set at 0.125 US dollars, to avoid effects associated with price discreteness; we control for prices not satisfying the arbitrage restriction in equation (15) in Bakshi et al. (1997). Similar filters have been applied in previous studies (Ait-Sahalia and Lo, 1998, Bakshi et al., 2000, and Christoffersen et al.,2006, among others). Finally, for each Wednesday, we consider options with maturities in maximum one year.

In performing the option pricing exercise, we consider S\&P500 index
levels recorded at 3 p.m. US Central Time (Chicago time), to maintain
the temporal alignment with the option prices (we consider the closing
price recorded at the close of the option trading, at 3 p.m. Chicago
time). Moreover, we follow Bakshi et al. (1997), Ait-Sahalia and Lo
(1998) and Christoffersen et al. (2006), and account for the effects
of dividends paid by the stocks included in the index. To perform the
pricing, we remove from the current index level the future dividends
which are expected to be paid during the life of the option. We
evaluate the dividends following Ait-Sahalia and Lo (1998), starting
from the relation between the index price and its future value:
\begin{equation*}
F_{t,\tau}=S_t e^{\left(r_{t,\tau}-d_{t,\tau}\right)\tau},
\end{equation*}
where $F_{t,\tau}$ is the Future price with maturity in $\tau$
periods, $r_{t,\tau}$ is the constant risk-free interest rate from $t$
to $t+\tau$ and $d_{t,\tau}$ is the constant, and unknown, divided
rate expected from $t$ to $t+\tau$. We determine the Future price from
the put-call parity
\begin{eqnarray}
P\left(K,t,\tau,r_{t,\tau}\right)+Ke^{-r_{t,\tau}\tau} &=& C\left(K,t,\tau,r_{t,\tau}\right)+S_t e^{-d_{t,\tau}\tau} \\
 &=& C\left(K,t,\tau,r_{t,\tau}\right)+F_{t,\tau} e^{-r_{t,\tau}\tau}. \label{formula:parity}
\end{eqnarray}
Note that the discounted Future price equals the Index level
discounted from future dividends. The dividend yield might be
recovered using the actual equity index level. In computing the
dividend discounted index, (\ref{formula:parity}) must be evaluated
using reliable option prices with, obviously, the same strike price
and the same maturity. We select, for each point in time (each
Wednesday) and each maturity, the pair of put and call options which
are closer to at-the-money (with the strike closest to the equity
index level). The interest rates are obtained from Thomson
Datastream. We download the interbank rates at 1, 3, 6 and 12
months. For options expiring in less than 41 (open market) days, we
use as risk-free the one month interbank rate. The three months rate
was used for values of $\tau$ between 41 and 82 days, while the six
months rate was considered for options expiring in more than 82 days
but less than 183 days. Finally, the 12 months rate was used for
options expiring in more than 183 days. After this procedure, for each
point in our sample and each maturity we obtain a set of discounted
index levels which are then used in the pricing, together with the set
of risk-free rates and the option prices previous described.


\begin{thebibliography}{999}
\bibitem{} Ait-Sahalia, Y., Lo, A.W., 1998, Nonparametric estimation in state-price densities implicit in financial asset prices. Journal of Finance 53, 499-547.
\bibitem{} Bakshi, G., Cao, C., Chen, Z., 1997, Empirical performance of alternative option pricing models. Journal of Finance 52, 2003-2050.
\bibitem{} Bakshi, G., Cao, C., Chen, Z., 2000, Pricing and hedging long-term options. Journal of Econometrics 94, 277-318.
\bibitem{} Baldovin, F., and Stella, A.L., 2007, 
Scaling and efficiency determine the irreversible evolution of a market,
Proc. Natl. Natl. Acad. Sci. USA 104, 19741-19744.
\bibitem{} Baldovin, F, Bovina, D., Camana, F., and Stella, A.L., 2011,
  Modeling the Non-Markovian, Non-stationary Scaling Dynamics of Financial Markets,
  in Abergel, F., Chakrabarti, B.K., Chakraborti, A. Mitra, M.
  (eds.), Econophysics of order-driven markets (1st edn), 
  New Economic Windows, Springer 239--252.
\bibitem{} Baldovin, F., Camana, F., Caporin, M., Caraglio, M., and Stella, A.L., 2014,
Ensemble properties of high frequency data and intraday trading rules,
Quantitative Finance.
\bibitem{} Ballocchi, G., Dacorogna, M.M., Gencay, R., and Piccinato, B., 1999, Intraday statistical properties of Eurofutures. Derivatives Quarterly 6, 28-44.
\bibitem{} Beran, J., 1994, Statistics for long-memory processes, Chapman \& Hall/CRC.
\bibitem{} Black F., Scholes M. (1973), The Pricing of Options and Corporate Liabilities. 
J. Polit. Econ. 81, 637-654.
\bibitem{} Borland, L., and Bouchaud, J.P., 2004, A non-Gaussian option pricing model with skew, Quantitative Finance, 4, 499-514.
\bibitem{} Bouchaud, J.P., Georges, A., 1990, Anomalous diffusion in disordered media: Statistical mechanisms, models and physical applications. Physics Reports 195, 127–293.
\bibitem{} Bouchaud, J.P., Potters M., 2003, Theory of Financial Risk and Derivative Pricing: from Statistical Physics to Risk Management, 2nd edn., Cambridge University Press.
\bibitem{} Chernov, M., Ghysels, E., 2000, A study toward a unified
  approach to the joint estimation of objective and risk neutral
  mearues for the purpose of option valuation. Journal of Financial
  Economics 56, 407-458.
\bibitem{} Christoffersen, P., Heston, S., Jacobs, K., 2006, Option
  valuation with conditional skewness. Journal of Econometrics 131,
  253-284.
\bibitem{} Christoffersen, P., Jacobs, K., 2004. Which GARCH model for
  option valuation? Management Science 50, 104-1221.
\bibitem{} Cont, R., 2001, Empirical properties of asset returns:
  stylized facts and statistical issues, Quantitative Finance, 1,
  223-236.
\bibitem{} Cont, R., 2005, Long range dependence in financial time series, 
in Lutton E., Levy V\'ehel J. (eds.), Fractals in Engineering, 
Springer-Verlag, New York.
\bibitem{} Corsi, F., 2009, A simple approximate long-memory model of realized volatility, Journal of Financial Econometrics 7, 174-196.
\bibitem{} Dacorogna, M.M., Muller U.A., Nagler, R.J., Olsen, R.B.,
  Pictet, O.V., 1993, A geographical model for the daily and weekly
  seasonal volatility in the foreign exchange market, Journal of
  International Money and Finance 12, 413-438.
\bibitem{} Dacorogna, M.M., Gencay, R., Muller U.A., Olsen, R.B., and
  Pictet, O.V., 2001, An introduction to high frequency
  finance. Academic Press, San Diego, CA.
\bibitem{} Di Matteo, T., Aste, T., and Dacorogna, M.M., 2005,
  Long-term memories of developed and emerging markets: using scaling
  analysis to characterize their stage of development. Journal of
  Banking and Finance 29, 827-851.
\bibitem{} Dumas, B., Fleming, J., Whaley, R., 1998, Implied
  volatility functions: empirical tests. Journal of Finance 53,
  2059-2106.
\bibitem{} Fouque, J.-P., Papanicolau, G., Sircar, K.R., 2000,
  Derivatives in Financial Markets with Stochastic Volatility,
  Cambridge University Press, New York.
\bibitem{} Francq, C., and Zakoian, J., 2010, GARCH Models. Structure,
  statistical inference and financial applications., Wiley, UK.
\bibitem{} Ghashghaie, S., Breymann, W., Peinke, J., Talkner, P., Dodge, Y., 1996,
  Turbulent cascades in foreign exchange markets,
  Nature 381, 767-770.
\bibitem{} Guillaume, D.M., Dacorogna, M.M., Dave, R.R., Muller, U.A., Olsen, R.B., Pictet, O.V., From the bird's eye to the microscope: a survey of the new stylized facts of the intra-daily foreign exchange market, Finance and Stochastics 1, 95-129.
\bibitem{} Hamilton, J.D., and Susmel, R., 1994,
  Autoregressive Conditional Heteroskedasticity and Changes in Regime,
  Journal of Econometrics 64, 307.
\bibitem{} Heston, S.L., 1993, A closed-form solution for options with stochastic volatility with applications to bond and currency options, Review of Financial Studies, 6, 337-343.
\bibitem{} Heston, S.L., Nandi, S., 2000, A closed-form GARCH option pricing model. Review of Financial Studies 13, 585-626.
\bibitem{} Hull, J.C., 2000, Options, Futures and Other Derivatives. Prentice-Hall.
\bibitem{} Hurst, H.E., 1951, Long-term storage capacity of reservoirs. Transactions of the American Society of Civil Engineers 116, 770-808.
\bibitem{} Lo, A.W., 1991, Long-term memory in stock market prices. Econometrica 59, 1279-1313.
\bibitem{} Mandelbrot, B.B., 1963, The variation of certain speculative prices. Journal of Business 36, 394-419.
\bibitem{} Mantegna, R.N., and Stanley, H.E., 1995, Scaling behaviour in the dynamics of an economic index. Nature 376, 46-49.
\bibitem{} Merton, R., 1973, Theory of Rational Option Pricing.
Bell Journal of Economics and Management Science 4 , 141-183.
\bibitem{} Muller, U.A., Dacorogna, M.M., Olsen, R.B., Pictet, O.V., Schwarz, M., and Morgenegg, C., 1990, Statistical study of foreign exchange rates, empirical evidence of a price change scaling law, and intraday analysis, Journal of Banking and Finance 14, 1189-1208.
\bibitem{} Peirano, P.P., and Challet, D., 2012,
  Baldovin-Stella stochastic volatility process and Wiener process
  mixtures,
  Eur. Phys. J. B 85, 276.
\bibitem{} Robinson, P.M., 2003, Time series with long-memory. Oxford University Press.
\bibitem{} Sethna, J.P., Dahmen, K.A., and Myers, C.R., 2001,
Crackling noise, Nature 410, 242-250. 
\bibitem{} Stanley, H.E., and Plerou, V., 2001, Scaling and universality in economics: empirical results and theoretical interpretation. Quantitative Finance 1, 563-567.
\bibitem{} Zamparo, M., Baldovin, F., Caraglio, M., and Stella, A.L., 2013,
  Scaling symmetry, renormalization, and time series modeling, Phys. Rev. E 88, 062808.
\end{thebibliography}
\end{document}